\documentclass[%
 reprint,
 superscriptaddress,
 floatfix,
 amsmath,amssymb,
 aps,
 prx,
floatfix,
]{revtex4-2}
\usepackage{enumerate}
\usepackage{graphicx}
\usepackage{caption, subcaption}
\usepackage{dcolumn}
\usepackage{bm}
\usepackage[colorlinks=true,linkcolor=blue,citecolor=blue,urlcolor=blue]{hyperref}

\usepackage{algorithm2e}
\RestyleAlgo{ruled} 

\usepackage{dsfont}

\usepackage{comment}
\usepackage{xcolor}

\begin{document}

\preprint{APS/123-QED}

\title{Avoiding barren plateaus in the variational determination of geometric entanglement}

\author{L. Zambrano}
\email{leonardo.zambrano@icfo.eu}
\affiliation{ICFO - Institut de Ciencies Fotoniques, The Barcelona Institute of Science and Technology, 08860 Castelldefels, Barcelona, Spain}

\author{A.~D.~Muñoz-Moller}%
\affiliation{Instituto Milenio de Investigaci\'on en \'Optica y Departamento de F\'isica, Facultad de Ciencias F\'isicas y Matem\'aticas, Universidad de Concepci\'on, Casilla 160-C, Concepci\'on, Chile}%

\author{M. Mu\~noz}
\affiliation{Departamento de Ingenier\'{\i}a Matem\'{a}tica, 
Facultad de Ciencias F\'{\i}sicas y Matem\'{a}ticas, Universidad de Concepci\'{o}n,
Casilla 160 C, Concepci\'{o}n, Chile}

\author{L. Pereira}
\affiliation{Instituto de F\'{\i}sica Fundamental IFF-CSIC, Calle Serrano 113b, Madrid 28006, Spain}

\author{A.~Delgado}
\affiliation{Instituto Milenio de Investigaci\'on en \'Optica y Departamento de F\'isica, Facultad de Ciencias F\'isicas y Matem\'aticas, Universidad de Concepci\'on, Casilla 160-C, Concepci\'on, Chile}

\date{\today}

\begin{abstract}
The barren plateau phenomenon is one of the main obstacles to implementing variational quantum algorithms in the current generation of quantum processors. Here, we introduce a method capable of avoiding the barren plateau phenomenon in the variational determination of the geometric measure of entanglement for a large number of qubits. The method is based on measuring compatible two-qubit local functions whose optimization allows for achieving a well-suited initial condition, from which a global function can be further optimized without encountering a barren plateau. We analytically demonstrate that the local functions can be efficiently estimated and optimized. Numerical simulations up to 18-qubit GHZ and W states demonstrate that the method converges to the exact value. In particular, the method allows for escaping from barren plateaus induced by hardware noise or global functions defined on high-dimensional systems. Numerical simulations with noise are in agreement with experiments carried out on IBM’s quantum processors for 7 qubits.
\end{abstract}

\maketitle

\section{\label{sec:1}Introduction}

The current generation of quantum processors has been described as noisy intermediate-scale quantum (NISQ) devices, which are characterized by a number of qubits in the hundreds, low gate fidelity, short coherence times, and limited qubit connectivity \cite{Preskill2018quantumcomputingin}. These features preclude the use of error-correcting techniques \cite{Krinner2022QEC,Chen2022QEC} and the implementation of deep quantum circuits, severely limiting its usability. Despite this adverse scenario, the family of variational quantum algorithms (VQAs) \cite{Cerezo2021VQA, Bharti2022VQA, Tilly2022VQA} emerges as a promising candidate to solve problems of practical interest in NISQ processors while offering advantages over classical computers. VQAs are hybrid quantum-classical algorithms, where a cost function is efficiently evaluated on a quantum processor and it is optimized using classical optimization routines. This approach is used by variational quantum eigensolvers  \cite{Peruzzo2014VQE}, quantum approximate optimization algorithms \cite{Farhi2014QAOA}, and quantum neural networks \cite{Biamonte2017, Beer2020, Abbas2021}, finding applications in quantum chemistry \cite{Lanyon2010chem, Hempel2018chem, Nam2020chem}, finances \cite{ Hong2014, Egger2020, Barkoutsos2020, Herman2022}, and quantum tomography \cite{Ferrie2014, Chapman2016, UtrerasAlarcn2019, Zambrano2020, Liu2020, Rambach2021, Xue2022}, among others. 

Recently, VQAs have been applied to estimate the entanglement of quantum states \cite{Wang2021, Wang2022}. Among them, the variational determination of geometric entanglement (VDGE) \cite{munoz2022variational} estimates the geometric measure of entanglement (GME) of pure multi-qubit quantum states $|\Psi\rangle$. The GME is an entanglement monotone given by the distance between $|\Psi\rangle$ and its nearest fully separable state $|\Phi\rangle$. It was first introduced for bipartite pure states \cite{SHIMONY1995} and later generalized to multipartite systems \cite{Barnum2001}. The GME has been used to study the distinguishability of multipartite quantum states using local operations \cite{Hayashi2006}, quantum phase transitions in spin models \cite{Orus2008}, Grover’s search algorithm \cite{Biham2002}, and entanglement witnesses \cite{Wei2003}. VDGE minimizes the infidelity between $|\Psi\rangle$ and a parameterized separable state $|\alpha (\boldsymbol{\theta})\rangle$ to calculate the GME. This procedure requires only local unitary operations on each qubit, so the implementation in a quantum processor is direct, requiring only circuits of depth 1. VDGE has been shown to be experimentally feasible, with experiments up to 5 qubits on IBM's quantum processors. 

Numerical simulations of VDGE for $25$-qubit GHZ and W states have been performed, showing the convergence of the algorithm even in such a high dimension \cite{munoz2022variational}. However, the simulations consider initial conditions close to the optimum. With random initial conditions, VDGE exhibits a clear lack of convergence. This is a common problem of current implementations of VQAs called the barren plateau (BP) phenomenon \cite{McClean2018}, which is characterized by lack of convergence due to a gradient of the cost function with a vanishing average and a standard deviation that decreases exponentially with the number of qubits by, for example, a global cost function \cite{Cerezo2021BP} or the presence of noise \cite{Wang2021noise}. The BP strongly limits the applicability of most VQAs, so the search for techniques to avoid BP becomes an important problem today. It has been shown that BPs can be avoided in some VQA by employing a local cost function instead of a global one \cite{Cerezo2021BP, Khatri2019, BravoPrieto2019, LaRose2019}, using better initial parameters for the optimization \cite{Grant2019initialization, dborin2022matrix}, detecting the barren plateaus with classical shadows \cite{sack2022avoiding}, reducing the expressiveness of the parametric circuit \cite{holmes2022connecting}, including mid-circuit measurements \cite{wiersema2021measurement}, and using quantum annealing \cite{mele2022avoiding} or geometric quantum machine learning \cite{Ragone2022}. Despite the advances on this topic, there is no general indication to avoid BP in a generic VQA, and moreover, it is not clear which technique will help to avoid BPs in VDGE. 

In this article, we propose a method to avoid BP in the context of VDGE. We call this improved VDGE (iVDGE) method, which complements VDGE by adding a previous stage aimed at finding better initial parameters in the search space so that VDGE does not fall into the BP. In this initial stage, the global infidelity is replaced by the average of two-qubit infidelities, that is, a local function, which is subsequently minimized. We show that this average bounds the infidelity, so minimizing it also minimizes the global infidelity. Furthermore, the value of these bounds can be efficiently obtained through an unbiased estimator, whose error decreases with the number of qubits. The usage of this local function instead of a global one drives the optimization out of the flat landscape associated with the BP with a higher probability than the standard VDGE. Our approach establishes an alternative strategy to avoid BP in the case of global functions and thus complements the arsenal of existing proposals \cite{Cerezo2021BP, Khatri2019, BravoPrieto2019, LaRose2019}. 

We carried out exhaustive numerical simulations to test the performance of iVDGE and compare it with VDGE. Simulations for random states up to 6 qubits show that iVDGE provides a speed-up of convergence and is more resource efficient than VDGE. Simulations for superpositions of GHZ and W states of 18 qubits show that while iVDGE converges near the optimum, VDGE exhibits BP. Noisy numerical simulations with 7 qubits for several resources show that BPs appear with a much smaller probability in iVDGE than in VDGE, being 10\% and 71\% respectively for a sample of 8192 shots. We experimentally test and corroborate our findings by running iVDGE and VDGE for a 7-qubit GHZ state on the IBM quantum processors \cite{IBMQ}.  

\section{Results}

\subsection{\label{sec:2} GME determination by VDGE}

The geometric measure of entanglement $E_\Psi$ of an $n$-qubit quantum state $|\Psi \rangle$ is defined through the optimization problem
\begin{align}\label{lambda_max}
	E_\Psi = \min_{|\Phi\rangle \in \text{Pro} \left(\mathcal{H}\right)} I_\Psi(\Phi),
\end{align}
where 
\begin{equation}\label{inf=gme}
	I_\Psi(\Phi) = 1-  | \langle \Psi |\Phi\rangle |^2
\end{equation}
is the infidelity between two $n$-qubit pure states and $\text{Pro} \left( \mathcal{H} \right) = \{|\Phi_1 \rangle \otimes \cdots \otimes |\Phi_n\rangle: |\Phi_j \rangle \in \mathcal{H}_{j} \}$ is the subset of pure product states of the Hilbert space $\mathcal{H} = \bigotimes_{j=1}^n \mathcal{H}_j$ of $n$ qubits. Thereby, $E_\Psi$ corresponds to the minimal infidelity between the state $|\Psi\rangle$ and the set $\text{Pro} \left( \mathcal{H} \right)$ \cite{Wei2003}. 

The variational determination of geometric entanglement (VDGE) attempts to find $E_\Psi$ by optimizing the infidelity over a variational ansatz. For this, we define the Hamiltonian 
\begin{align}
    H_G = \mathds{1} - |0 \rangle \langle 0 |^{\otimes n}
\end{align}
and the variational ansatz
\begin{align}
    |\alpha (\boldsymbol{\theta})\rangle = U^\dagger (\boldsymbol{\theta}) |\Psi \rangle.
\end{align}
Here $U (\boldsymbol{\theta}) =  \bigotimes_{j = 1}^n U_j (\boldsymbol{\theta}_j)$, where $U_j (\boldsymbol{\theta}_j)$ is parametric single-qubit unitary operation acting on the $j$th qubit.
Thus, the cost function
\begin{align}\label{eq:gcost_function}
    I_\Psi(\boldsymbol{\theta}) = 1 - \langle \Psi | U (\boldsymbol{\theta}) |0 \rangle \langle 0 |^{\otimes n}  U^\dagger (\boldsymbol{\theta}) | \Psi \rangle
\end{align}
now depends on the parameters of ansatz and can be cast as
\begin{equation}
I_\Psi(\boldsymbol{\theta})=\langle H_G \rangle_{\boldsymbol{\theta}} = \langle \alpha (\boldsymbol{\theta}) | H_G |\alpha (\boldsymbol{\theta})\rangle.
\end{equation}
This can be estimated by repeatedly preparing the $n$-qubit state $|\Psi \rangle$, applying the circuit that implements $U (\boldsymbol{\theta})$, and measuring in the computational basis.


The infidelity $I_\Psi(\boldsymbol{\theta})$ is minimized by an optimization method running on a classical computer. Typical choices for the optimization method are the Simultaneous Perturbation Stochastic Approximation (SPSA) method \cite{spallspsa1992} and, more recently, its complex version CSPSA \cite{UtrerasAlarcn2019}. These methods require in each iteration the estimation of $I_\Psi(\boldsymbol{\theta})$ in two different points of the optimization space. Therefore, after a prescribed number of iterations, VDGE provides an estimate $\hat E_\Psi$ of GME. The minimization of GME is, however, a challenging problem since the optimization landscape might exhibit many local minima. This is usually overcome by repeating the optimization several times with different initial conditions to obtain a set of estimates $\{\hat{E}_\Psi^k \}$. The final estimate is given by the minimum over this set, that is, $\hat{E}_\Psi=\min_k\{\hat{E}_\Psi^k \}$.

Numerical simulations of VDGE for 25-qubit GHZ and W states exhibit the barren plateau phenomenon \cite{munoz2022variational}. This occurs because for a $n$-qubit system both the infidelity $I_\Psi(\boldsymbol{\theta})$ and its gradient vanish exponentially with the number of qubits when the initial parameters are randomly chosen \cite{Most_Gross}. One alternative to overcome this problem is to have a better initial condition, which can be obtained, for instance,  by resorting to a priori information \cite{Grant2019initialization, dborin2022matrix} about the solution. Another alternative is to overcome the sampling noise by employing more than $O(2^n)$ measurement shots, which leads to accurate evaluations of $I_\Psi(\boldsymbol{\theta})$ and its gradient. This, however, does not guarantee to avoid the BP, and moreover, the exponential scaling of the shot number makes the algorithm unfeasible.


\subsection{\label{sec:3} Improved VDGE}
To prevent the BP phenomenon from harming the performance of VDGE we resort to a local function instead of the global infidelity Eq.~\eqref{eq:gcost_function} \cite{Cerezo2021VQA}. We define this function as the expectation value of the local Hamiltonian 
\begin{equation}\label{eq: local hamiltonian}
    H_L = \mathds{1} - \frac{2}{n(n-1)} \sum_{\substack{i, j =1 \\ i < j}}^n \Pi_{i j},
\end{equation}
where each operator $\Pi_{i j}$ acts as a rank-1 projector on qubits $i$ and $j$, that is, 
\begin{align}\label{eq: local projectors}
    \Pi_{i j} =  |0_i \rangle \langle 0_{i}| \otimes |0_j \rangle \langle 0_j|,
\end{align} 
and as an identity operator on any other qubit. The Hamiltonian $H_L$ is constructed such that $| 0\rangle^{\otimes n}$ is its ground state and its expectation value is given by the average of the two-qubit infidelities
\begin{align}\label{opt_local}
        I_{ij}(\boldsymbol{\theta}_i, \boldsymbol{\theta}_j)=1 - \mathrm{Tr} \left(  \rho_{i,j} (\boldsymbol{\theta}_i, \boldsymbol{\theta}_j)  \; \Pi_{ij} \right),
\end{align}
where $\rho_{i,j}$ is the reduced density matrix of qubits $i$ and $j$ from state $|\alpha (\boldsymbol{\theta})\rangle$.

The expectation value $\langle H_L \rangle_{\boldsymbol{\theta}} = \langle \alpha (\boldsymbol{\theta}) | H_L |\alpha (\boldsymbol{\theta})\rangle $ can be used to obtain the following upper and lower bounds of the infidelity (see Appendix \ref{sec: Method} for details)
\begin{align}\label{eq: Infidelity bounds}
\langle H_L \rangle_{\boldsymbol{\theta}}  \leq I_\Psi (\boldsymbol{\theta})  \leq \left\lfloor \frac{n}{2} \right\rfloor \langle H_L \rangle_{\boldsymbol{\theta}},
\end{align}
which are tight if and only if $| \Psi \rangle$ is a product state. According to Eq.~(\ref{eq: Infidelity bounds}) it is possible to obtain a first estimate of the GME by minimizing the local function $\langle H_L \rangle_{\boldsymbol{\theta}}$ instead of the global function $\langle H_G \rangle_{\boldsymbol{\theta}}$. 

The expectation value of the Hamiltonian $H_L$ can be cast in the form
\begin{equation}
    \langle H_L \rangle_{\boldsymbol{\theta}} = \frac{1}{\vert \mathcal{G} \vert} \sum_{g\in\mathcal{G}} X_g,
\end{equation}
where
\begin{equation}
    X_g = \frac{1}{\lfloor n/2 \rfloor}\sum_{(i,j)\in g} I_{ij} (\boldsymbol{\theta}_i,\boldsymbol{\theta}_j)
\end{equation}
and the set $\mathcal{G}$ contains all groups $g$ of partitions of $n$ qubits formed by non-overlapping pairs of qubits. This suggests that the expectation value $\langle H_L\rangle_{\boldsymbol{\theta}}$ can be evaluated by sampling $X_g$ uniformly on the set $\mathcal{G}$. For a given group $g$ the evaluation of the infidelities in $X_g$ can be performed efficiently since the measurements are local and can be carried out in parallel. To keep resource usage to a minimum, we randomly select a single $X_g$ as an estimator of $\langle H_L \rangle_{\boldsymbol{\theta}}$. This estimator is unbiased, that is, $\mathbb{E}(X_g)=\langle H_L\rangle_{\boldsymbol{\theta}}$, and its mean square error (MSE) is bounded and decreases inversely proportional with the number of qubits (see Appendix \ref{sec: Sampling} for details),
\begin{equation}
    \textrm{MSE}\left(X_g \right) \leq  \frac{n-1}{(n-2)(n-3)} 
\end{equation}
The algorithm that we propose, which we call improved VDGE (iVDGE), arises from concatenating the following two stages: (i) the minimization of the expectation value of the local Hamiltonian $H_L$ to obtain a first estimate of the GME, and (ii) the execution of a standard VDGE, that minimizes the expectation value of the global Hamiltonian $H_G$, but using as initial guess the estimator obtained from the stage (i). The concatenation of these steps allows us to avoid BPs and accurately estimate the GME.

Stage (i) is performed as follows:

\begin{enumerate}[$(i)$]
    \item[(a)] Split the $n$ qubits in a partition of $\lfloor n/2 \rfloor $ pairs $(i, j)$ at random, plus a single qubit if $n$ is odd, or equivalently, sample a single group $g\in\mathcal{G}$ of non-overlapping pairs of qubits.
    
    \item[(b)] For each pair $(i, j)$ in the partition, perform a single iteration of a classical optimization method to minimize the corresponding local infidelity \eqref{opt_local}. This is equivalent to minimizing the expected value of the local Hamiltonian $H_L$.
    
    \item[(c)] Repeat steps $(a)$ and  $(b)$ a predefined number $N_L$ of times according to a classical optimization method. 
\end{enumerate}

Notice that the selection of the classical optimization method is fundamental to having a good performance with
iVDGE. Particularly, we chose the CSPSA method~\cite{UtrerasAlarcn2019} (see Appendix \ref{sec:CSPSA}). This is a stochastic optimization algorithm that works over complex parameters. Due to that, in our case $\boldsymbol{\theta}$ is a complex vector of $2n$ elements. In algorithm \ref{alg:iVDGE} we present the pseudocode of the iVDGE when CSPSA is used as classical optimization method.

The iVDGE method might still get trapped in local optima. To avoid them, we run the optimization several times starting from different separable initial states to obtain a set of possible estimators of the GME, and keep the best value obtained as the final estimator.

\begin{algorithm}[t!]
\caption{iVDGE algorithm}
\label{alg:iVDGE}    
Input: Quantum state $\Psi$, number of local iterations $N_L$, number of global iterations $N_G$\\
Initialization: parameters $\theta_0 \in \mathbb{C}^{2\times n}$ \\
\For{$k=1,\ldots, N_L+N_G$}{
    \eIf{$k\leq N_L$}{
   Set a partition $\left\{\Omega_i\right\}_{i=1}^{\lfloor n/2\rfloor}$   of $\{1,n\}$ \\
  \For{$i=1,\ldots ,\lfloor n/2\rfloor $}{
    Compute CSPSA gradient $\nabla I_{\Psi,\Omega_i}(\theta_{k-1})$\\ 
    $\theta_k = \theta_{k-1} - a_k\nabla I_{\Psi,\Omega_i}(\theta_{k-1})$
    }}{
    Compute CSPSA gradient $\nabla I_{\Psi}(\theta_{k-1})$\\ 
    $\theta_k = \theta_{k-1} - a_k\nabla I_{\Psi}(\theta_{k-1})$
    }
    }
\end{algorithm}

\begin{figure*}
	\centering
	\begin{subfigure}[b]{0.45\textwidth}
		\includegraphics[width=\textwidth]{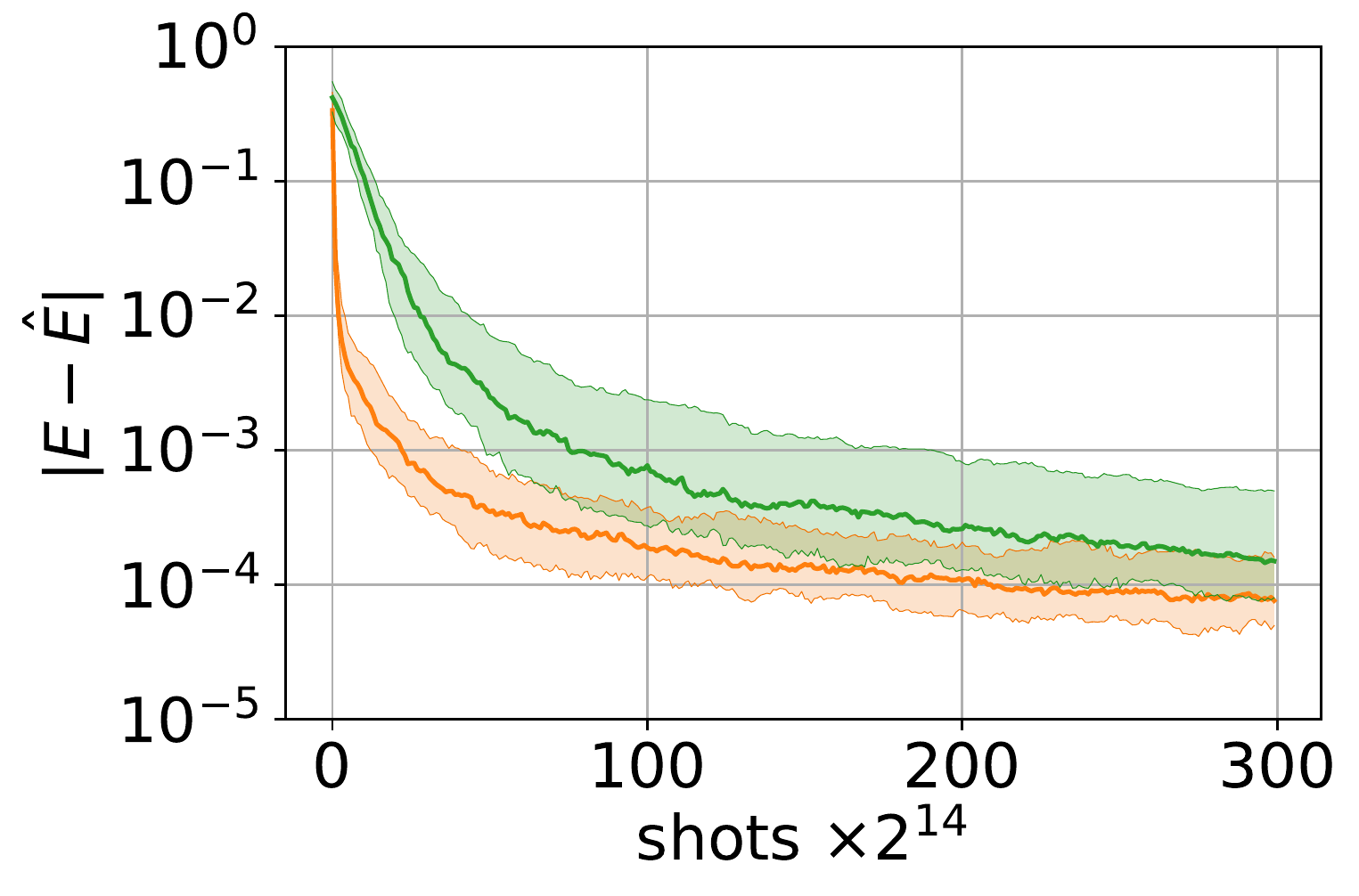}
		\caption{$3$ qubits.}
		\label{fig:Fig1-1}
    \end{subfigure}
	\begin{subfigure}[b]{0.45\textwidth}
		\includegraphics[width=\textwidth]{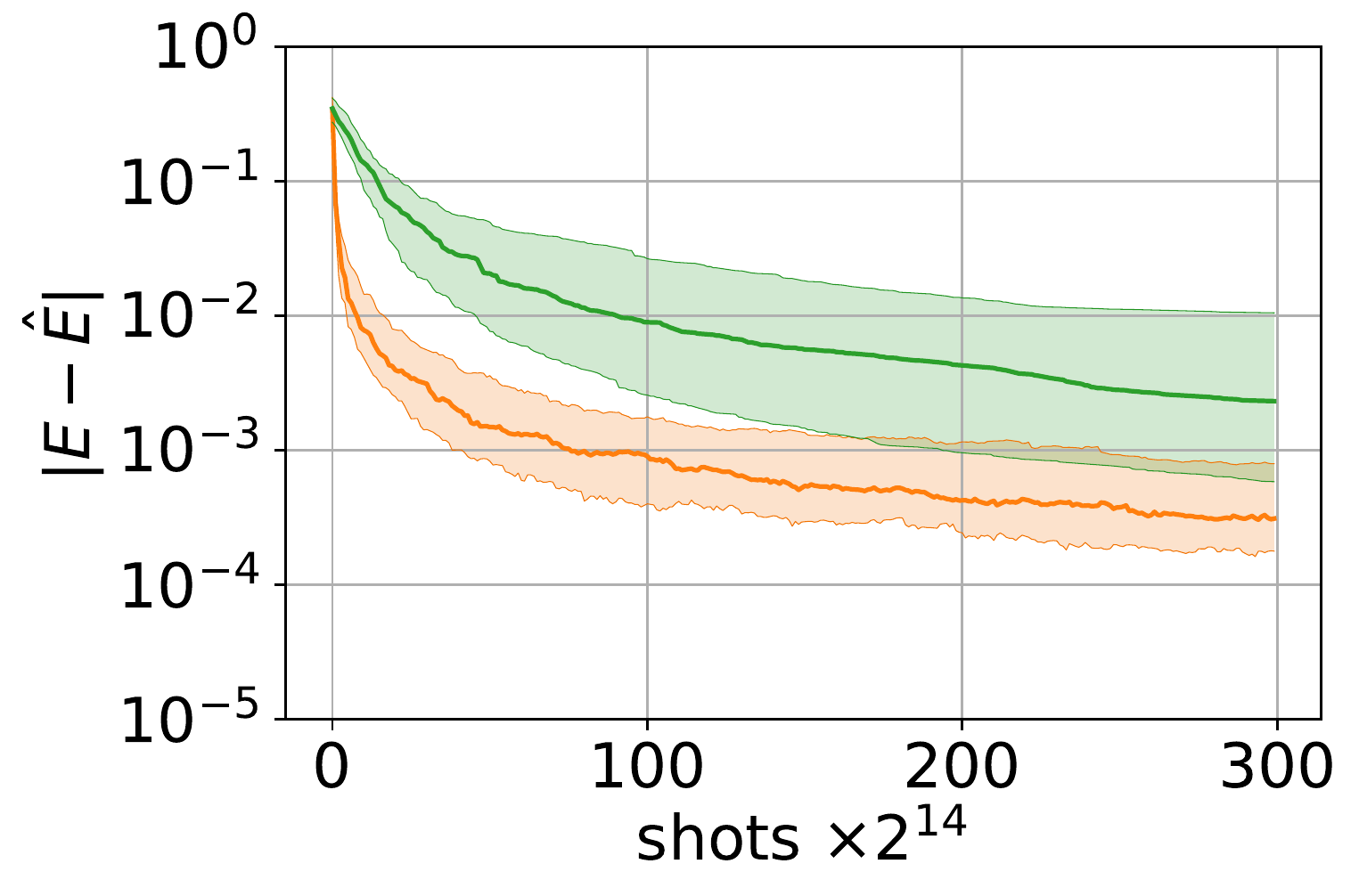}
		\caption{$4$ qubits. }
		\label{fig:Fig1-2}
	\end{subfigure}
	\begin{subfigure}[b]{0.45\textwidth}
		\includegraphics[width=\textwidth]{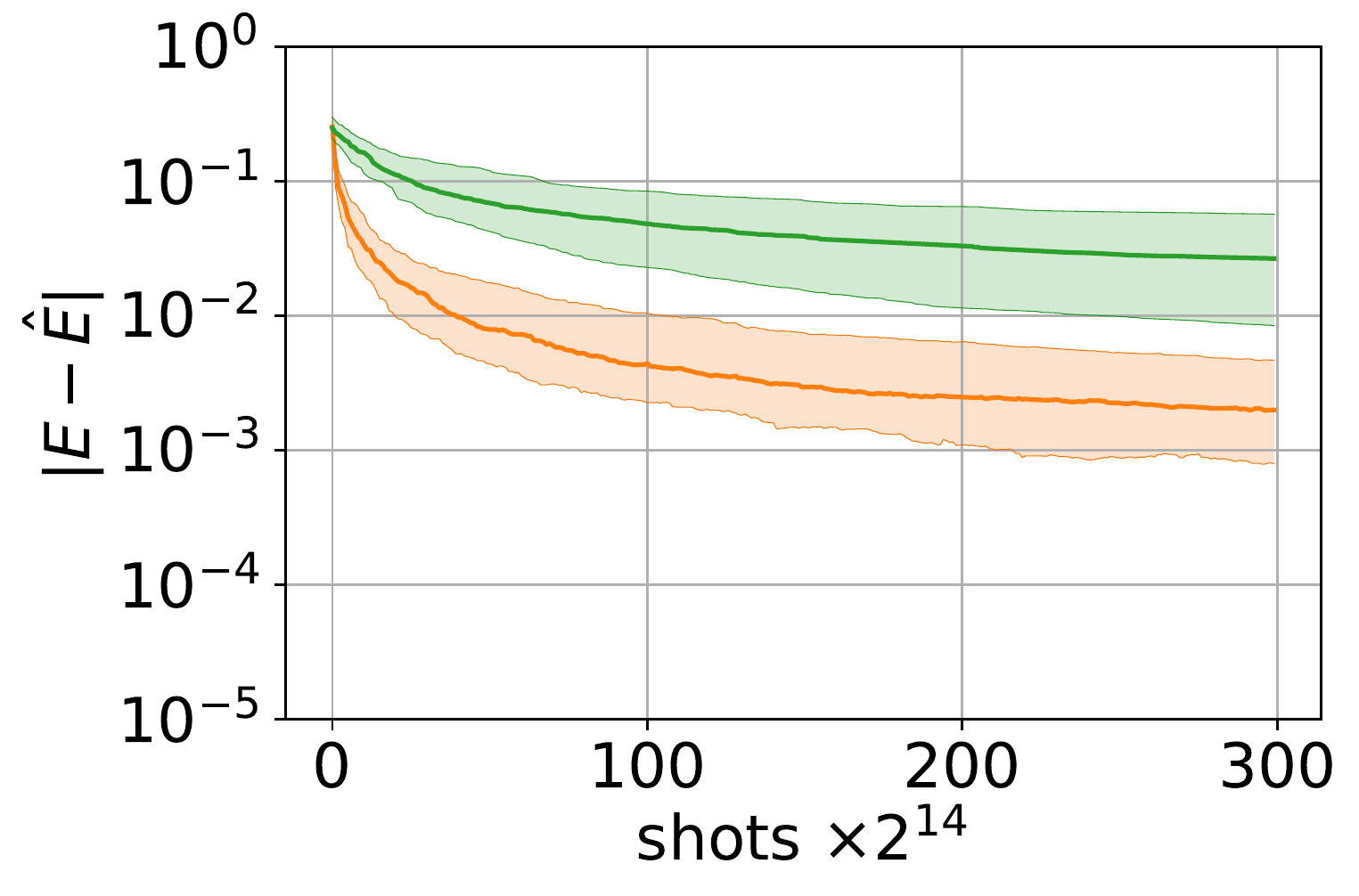}
		\caption{$5$ qubits.}
		\label{fig:Fig1-3}
	\end{subfigure}
	\begin{subfigure}[b]{0.45\textwidth}
		\includegraphics[width=\textwidth]{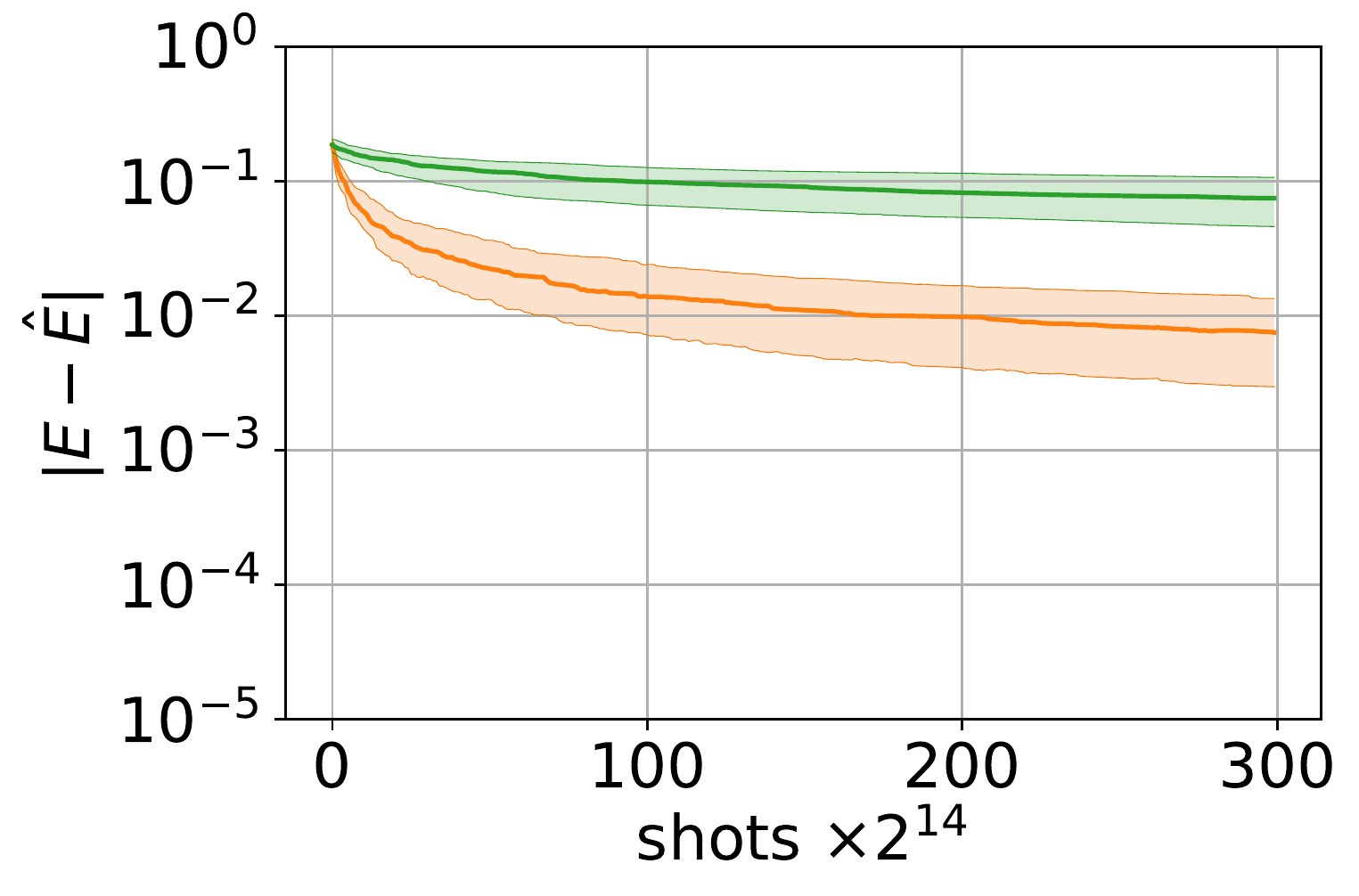}
		\caption{$6$ qubits.}
		\label{fig:Fig1-4}
    \end{subfigure}
    \caption{Difference $|E-\hat{E}|$ between the exact GME $E$ and the estimated GME $\hat{E}$ values obtained with VDGE (green) and iVDGE (orange) as function of the number of shots. Solid lines indicate the median difference calculated over an ensemble of 100 randomly selected pure states. Shaded areas represent the corresponding interquartile range.}
\label{fig:Fig1}
\end{figure*}

\subsection{\label{sec:4}Numerical Simulations}
We carried out numerical simulations of iVDGE and VDGE in order to test and compare their performances. 

In our first test, we consider simulations with random initial conditions for a small number of qubits. We generate a set of $100$ pure states that are randomly chosen according to a Haar-uniform distribution for $3$, $4$, $5$, and $6$ qubits. For each state, we estimate the GME by both methods. For iVDGE, the first stage is implemented with $80$ iterations, using $512$ shots to simulate each measurement of the infidelity. The second stage is executed with $295$ iterations and using $8192$ shots per infidelity evaluation. The first stage uses a total of shots equal to that used in five iterations of the second stage. On the other hand, VDGE is implemented using $300$ iterations with $8192$ shots per infidelity evaluation. Thereby, our implementations of iVDGE and VDGE use the same total amount of shots. To avoid local optima, this procedure is repeated $5$ times for each randomly generated state in both methods and the minimum GME value obtained is the final GME estimate. We also numerically compute the exact value of the GME for each of the randomly generated states through the Basin-hopping optimization algorithm \cite{bashop} (see Appendix \ref{sec:A}).

\begin{figure*}[t!]
	\centering
	\begin{subfigure}[b]{0.45\textwidth}
    	\centering
    	\includegraphics[scale=0.55]{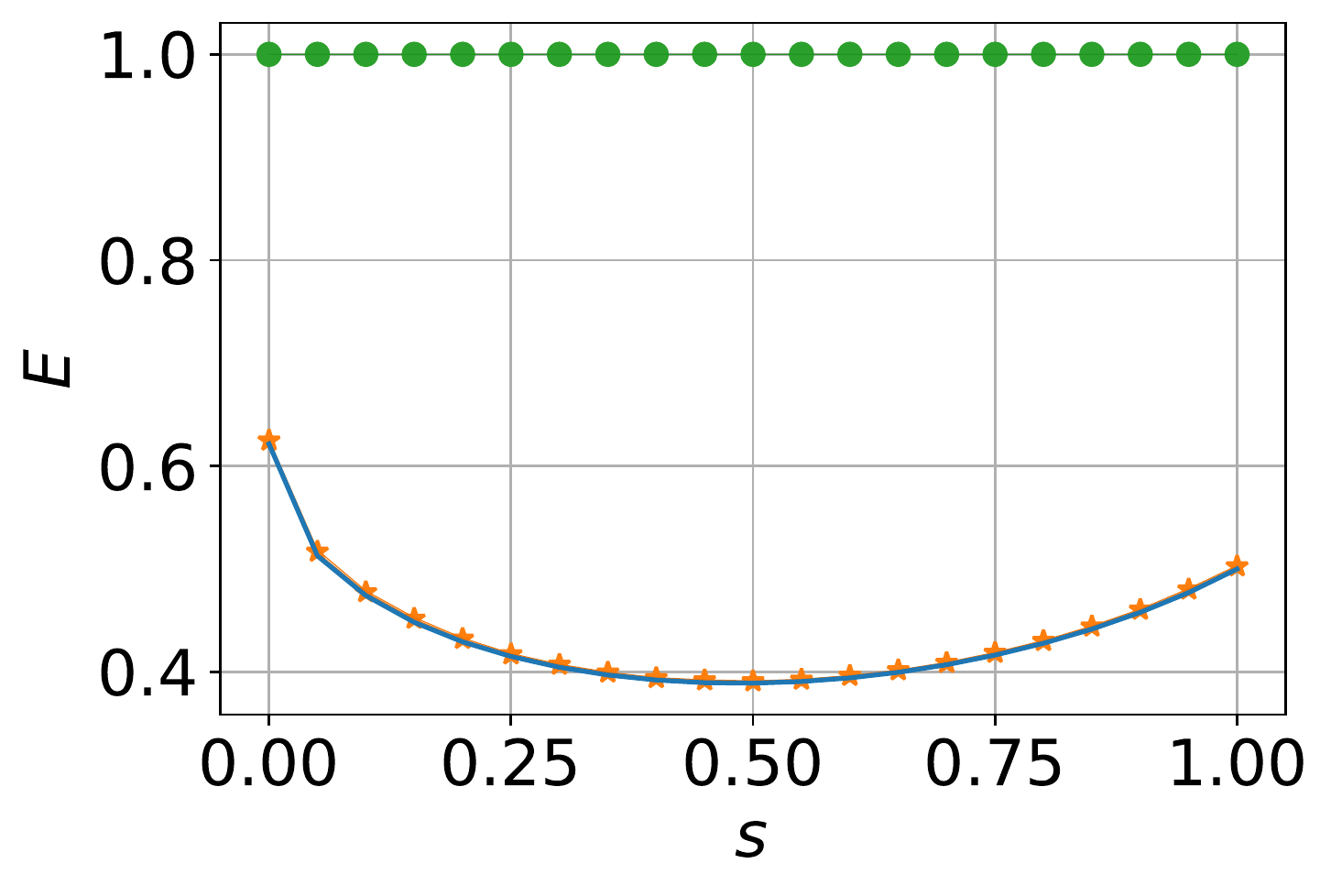}
    	\caption{GME for $|\text{GHZW} (s) \rangle$.}
    	\label{fig:Fig2-1}
    \end{subfigure}
	\begin{subfigure}[b]{0.45\textwidth}
    	\centering
    	\includegraphics[scale=0.55]{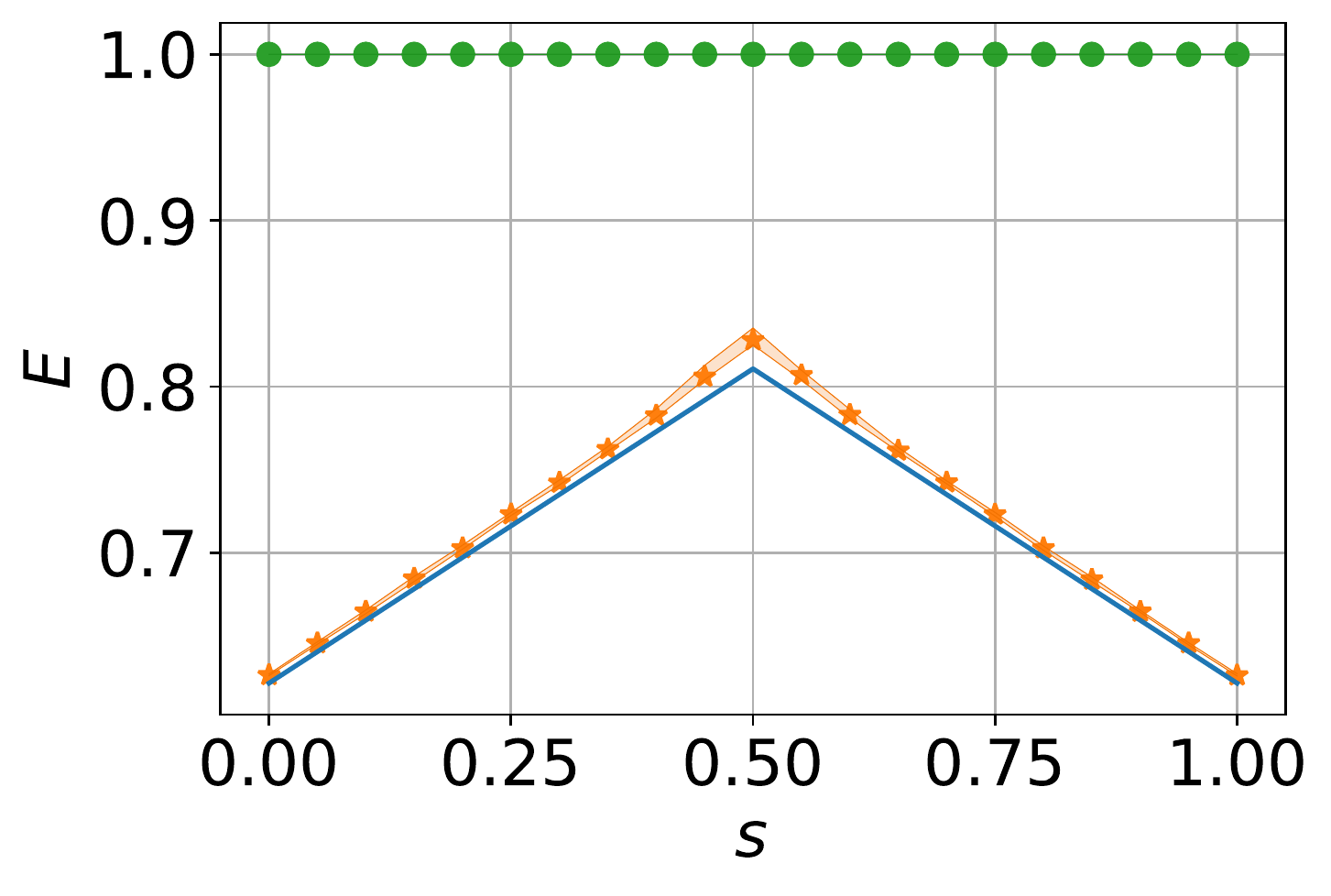}
    	\caption{GME for $|\text{W\~W} (s) \rangle$.}
    	\label{fig:Fig2-2}
	\end{subfigure}
	\caption{Value of the GME for 18-qubit $|\text{GHZW} (s) \rangle$ and $|\text{W\~W} (s) \rangle$ states as a function of $s$. Orange stars (green dots) represent the median values obtained using iVDGE (VDGE) on a set of 100 repetitions with randomly selected initial conditions. The blue curve is the exact GME value. Shaded areas correspond to interquartile ranges.}
\label{fig:Fig2}
\end{figure*}
\begin{figure}[t!]
	\centering
	\includegraphics[scale=0.55]{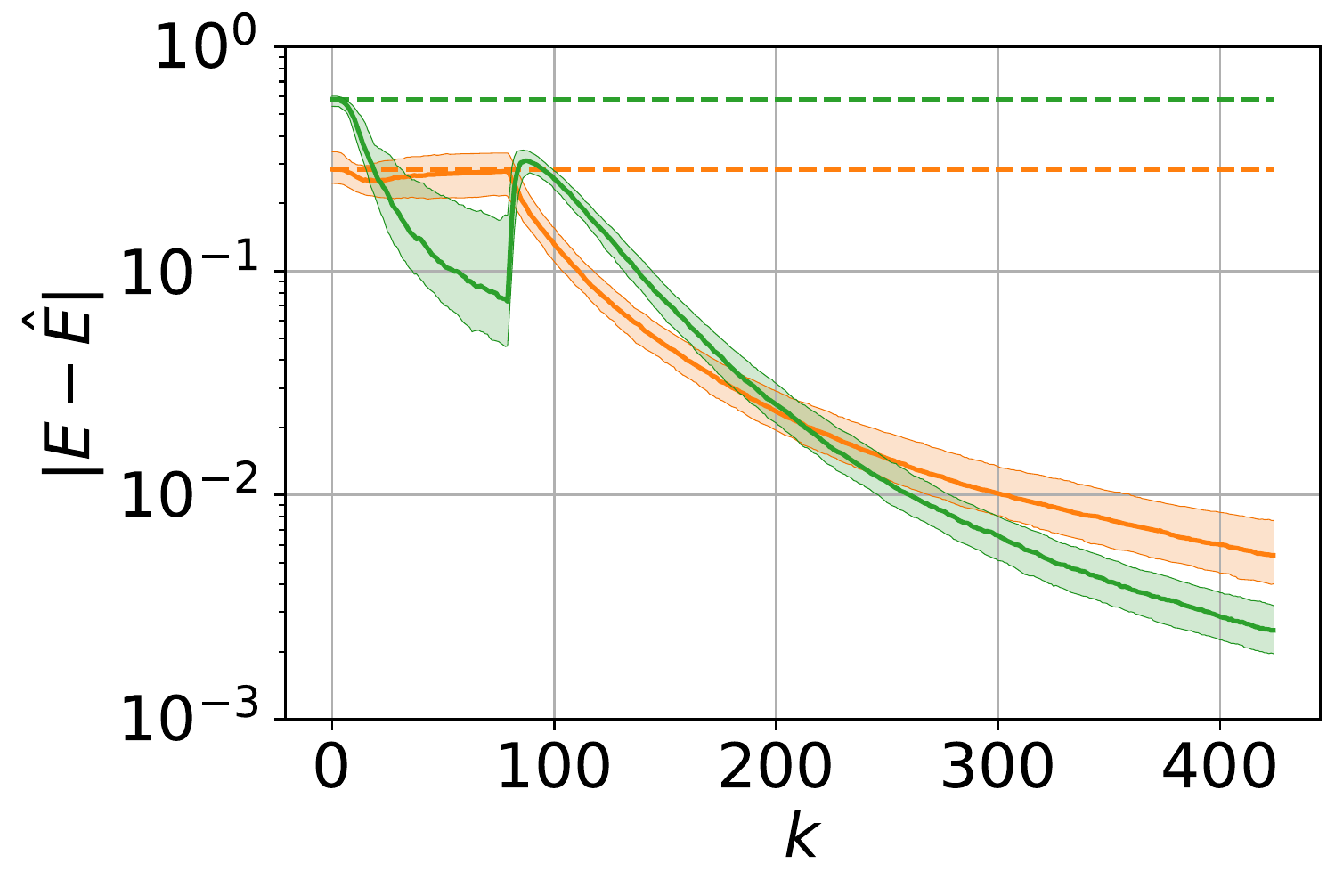}
	\caption{Median of the difference $|E-\hat{E}|$ between the exact GME $E$ and the estimated GME $\hat{E}$ values on the sets of 18-qubit $|\text{GHZW} (s) \rangle$ (green) and $|\text{W\~W} (s) \rangle$ (orange) states as a function of the number of iterations $k$. The median is calculated from the set of 100 GME estimates for all 21 values of $s$ using iVDGE. The first $80$ iterations correspond to the optimization of local functions and the next $345$ to the optimization of the global function. Dotted lines correspond to the median of the difference $|E-\hat{E}|$ obtained via VDGE. Shaded areas correspond to interquartile ranges.}
\label{fig:Fig3}
\end{figure}

The comparison between the performances of iVDGE and VDGE is shown in Fig.~\ref{fig:Fig1}. The orange (green) curve represents the median of the difference $|E - \hat{E}|$ between $E$, the exact GME value, and $\hat{E}$, the values obtained through iVDGE (VDGE) as a function of the number of accumulated shots. Shaded areas correspond to interquartile ranges. As Fig.~\ref{fig:Fig1} indicates, iVDGE achieves faster convergence and higher accuracy than VDGE in all four cases. As the number of qubits increases, the interquartile ranges of iVDGE and VDGE do not intersect and the estimation precision of GME decreases and presents flattening, the latter being less pronounced for the iVDGE method.

To comparatively study the performance of iVDGE and VDGE in the high qubit number regime, where the BP phenomenon is expected to arise, we perform simulations with particular quantum states, since the calculation of the GME for arbitrary states in this regime becomes unfeasible \cite{Most_Gross}. We resort to the GHZ, W, and \~W states for $n$ qubits given by the expressions.
\begin{align}
    | \text{GHZ} \rangle &= \frac{1}{\sqrt{2}}( |00 \cdots 0 \rangle + |11 \cdots 1\rangle), \\
    |\text{W} \rangle &= \frac{1}{\sqrt{n}}( | 10\cdots0 \rangle + | 01\cdots0 \rangle +\cdots +|0\cdots01\rangle, \\
    |\text{\~W} \rangle &= \frac{1}{\sqrt{n}}( | 01\cdots1 \rangle + | 10\cdots1 \rangle +\cdots +|1\cdots10\rangle.
\end{align}
These states have GME values of $0.5$ for the GHZ state and $1 - [(n-1)/n)]^{n-1}$ for W and \~W states \cite{Wei2003}.  For superpositions of the form
\begin{align}
    |\text{GHZW} (s) \rangle = \sqrt{s} | \text{GHZ}\rangle +  \sqrt{1 - s}| \text{W} \rangle,
\end{align}
or 
\begin{align}
    |\text{W\~W} (s) \rangle = \sqrt{s} | \text{W}\rangle +  \sqrt{1 - s}| \text{\~W} \rangle,
\end{align}
with $s \in [0, 1]$, the GME is optimized by a symmetric separable state \cite{suma_ghz_w}, that is,  $|\Phi \rangle \otimes \dots \otimes |\Phi \rangle$. This reduces the optimization space to only two parameters, making it easier to evaluate the exact value of the GME using classical optimization techniques.

Fig.~\ref{fig:Fig2} exhibits the value of the GME obtained with VDGE and iVDGE for the states $ |\text{GHZW} (s) \rangle$ and $|\text{W\~W} (s) \rangle$ as a function of $s$ and for $n=18$ qubits. For each state and value of $s = 0.05m$ with $m = 0, 1, \dots, 20$, VDGE and iVDGE are executed with 5 randomly selected initial conditions and the minimum value of GME is selected as an estimate. This procedure is repeated 100 times. Orange stars represent the median value obtained using iVDGE with $80$ iterations in the first stage, and $295$ iterations in the second stage. The infidelities were approximated with $512$ and $8192$ shots, respectively. Green circles represent the median value obtained using VDGE with $300$ iterations, evaluating the infidelity with $8192$ shots. The solid blue curve represents the exact value of the GME (see Appendix \ref{sec:A}). Shaded areas correspond to interquartile ranges, which are extremely narrow. We see that even with such scarce resources, iVDGE converges to the GME, while VDGE never escapes a value other than one. This suggests that the BP phenomenon is present in VDGE, while iVDGE is able to avoid it, converging to the exact value.

Fig.~\ref{fig:Fig3} allows us to analyze the convergence of iVDGE from a different perspective using the data from the previous simulation. This figure displays the median error $|E - \hat{E}|$ calculated on the set of 100 GME estimates for all 21 values of $s$ for $|\text{GHZW} (s) \rangle$ states (green) and $|\text{W\~W} (s) \rangle$ states (orange), as a function of the number of iterations. Solid and dashed lines display the median of iVDGE and VDGE, respectively. Shaded areas correspond to interquartile ranges. VDGE and iVDGE start at the same point in the parameter space. In the first $80$ iterations, iVDGE tries to find a good starting point for the global optimization at the second stage. For the orange curve, we see that although the distance from the optimum does not improve significantly in the first stage, at the beginning of the second stage iVDGE is effectively outside the BP and also converges. For the green curve, we see that the distance from the optimum increases when iVDGE switches to the second stage. This behaviour might be controlled by fine-tuning the hyper-parameters of the optimization algorithm. Nevertheless, iVDGE does not enter the BP again and converges with good accuracy.

To study the ability of the iVDGE algorithm to bypass noise-induced barren plateaus \cite{Wang2021noise}, we performed noisy simulations for both VDGE and iVDGE using limited resources and the noise model of IBM Quantum system \texttt{ibm\_oslo}, which includes one- and two-qubit gate errors as well as measurement errors. The geometric measure of entanglement of a 7-qubit GHZ state is calculated for 100 randomly generated initial conditions with VDGE and iVDGE, using different numbers of shots. For the VDGE algorithm, we performed 200 iterations, while for the iVDGE algorithm, we performed 80 local iterations plus the required amount of global iterations such that the same total number of shots are used in both methods. For example, to match the 200 iterations of VDGE with 8192 shots, iVDGE uses 80 local iterations with 512 shots and 195 global iterations with 8192 shots. On the last iteration, we perform error mitigation using matrix inversion to mitigate the measurement error. Unlike the previous simulations, we do not consider extra repetitions with random initial conditions to avoid local optima. The GME of a GHZ is equal to $0.5$, but the state prepared by the circuit has a GME of $0.49177$ due to noisy operations that decrease the amount of entanglement. We consider that the algorithm is trapped in a BP if, after the optimization, the GME is still greater than $0.9$.

Tables \ref{BP_VDGE} and \ref{BP_iVDGE} show the percentage of BPs appearance for a given amount of total shots for the VDGE and iVDGE algorithms, respectively, together with the median estimated geometric measure of entanglement for the 100 different initial conditions. The simulations show that in the case of VDGE, the BP appearance is approximately independent of the number of shots, where about $70\%$ of the realizations
exhibit the BP phenomenon. On the other hand, the BP occurrence for the iVDGE simulations stays relatively low using 512 local shots and up to 1024 global shots but increases rapidly by further reducing the total number of shots. The median GME only changes drastically when using the lowest amount of shots. The iVDGE results not only show a better convergence in comparison to the VDGE method but also a better median GME for every set of shots.

\begin{table}[]
\begin{tabular}{|ccc|}
\hline
\multicolumn{3}{|c|}{VDGE}                                                   \\ \hline
\multicolumn{1}{|c|}{Global shots} & \multicolumn{1}{c|}{BP \%} & Median GME \\ \hline
\multicolumn{1}{|c|}{8192}         & \multicolumn{1}{c|}{71\%}  & 0.9822     \\ \hline
\multicolumn{1}{|c|}{4096}         & \multicolumn{1}{c|}{69\%}  & 0.9855     \\ \hline
\multicolumn{1}{|c|}{2048}         & \multicolumn{1}{c|}{74\%}  & 0.9830     \\ \hline
\multicolumn{1}{|c|}{1024}         & \multicolumn{1}{c|}{69\%}  & 0.9854     \\ \hline
\multicolumn{1}{|c|}{512}          & \multicolumn{1}{c|}{74\%}  & 0.9836     \\ \hline
\multicolumn{1}{|c|}{256}          & \multicolumn{1}{c|}{73\%}  & 0.9866     \\ \hline
\multicolumn{1}{|c|}{130}          & \multicolumn{1}{c|}{77\%}  & 0.9828     \\ \hline
\end{tabular}
\caption{Percentage of BPs appearances and median GME value of a 7-qubit GHZ state with 100 randomly generated initial conditions using the VDGE algorithm with different numbers of shots. For every case, 200 iterations are considered.}
\label{BP_VDGE}
\end{table}

\begin{table}[]
\begin{tabular}{|cccc|}
\hline
\multicolumn{4}{|c|}{iVDGE}                                                                                     \\ \hline
\multicolumn{1}{|c|}{Local shots} & \multicolumn{1}{c|}{Global shots} & \multicolumn{1}{c|}{BP \%} & Median GME \\ \hline
\multicolumn{1}{|c|}{512}         & \multicolumn{1}{c|}{8192}         & \multicolumn{1}{c|}{10\%}  & 0.5322     \\ \hline
\multicolumn{1}{|c|}{512}         & \multicolumn{1}{c|}{4096}         & \multicolumn{1}{c|}{10\%}  & 0.5322     \\ \hline
\multicolumn{1}{|c|}{512}         & \multicolumn{1}{c|}{2048}         & \multicolumn{1}{c|}{9\%}   & 0.5331     \\ \hline
\multicolumn{1}{|c|}{512}         & \multicolumn{1}{c|}{1024}         & \multicolumn{1}{c|}{9\%}   & 0.5407     \\ \hline
\multicolumn{1}{|c|}{256}         & \multicolumn{1}{c|}{512}          & \multicolumn{1}{c|}{16\%}  & 0.5516     \\ \hline
\multicolumn{1}{|c|}{128}         & \multicolumn{1}{c|}{256}          & \multicolumn{1}{c|}{20\%}  & 0.5951     \\ \hline
\multicolumn{1}{|c|}{64}          & \multicolumn{1}{c|}{128}          & \multicolumn{1}{c|}{39\%}  & 0.7698     \\ \hline
\end{tabular}
\caption{Percentage of BPs appearances and median GME value of a 7-qubit GHZ state with 100 randomly generated initial conditions using the iVDGE algorithm with different numbers of shots. For every case, 80 iterations are performed in the local stage, while the number of iterations in the global stage is chosen such that the same total number of shots are used in both methods.}
\label{BP_iVDGE}
\end{table}

\subsection{\label{sec:5}Experimental results for 7-qubit GHZ state}

We perform an experimental demonstration of our method using the IBM quantum processor \texttt{ibm\_oslo}. We measure the GME of a $7$-qubit GHZ state using iVDGE and VDGE, repeating the procedure 12 times with different initial conditions. For iVDGE, we perform $80$ local and $260$ global iterations using $128$ and $256$ shots, respectively, to estimate the infidelities. For VDGE we perform $300$ iterations using $256$ shots for each measurement of the infidelity. As the performance of the quantum device is affected by various sources of noise, we can not generate an ideal GHZ state. Because of this, we perform quantum state tomography using $20000$ shots per basis with maximum likelihood estimation \cite{Hradil1997_MLE, Shang2017_superfast} using a pure state parameterization and error mitigation \cite{Bravyi2021_errormitigation}, obtaining an estimated GHZ state that has fidelity of $0.91 \pm 0.01$ with respect to the ideal GHZ. The GME of the estimated state is obtained using the Basin-hopping optimization algorithm, which leads to a GME value of $0.519 \pm 0.01$.

To estimate the experimental errors of both the fidelity and the GME, we sample 100 times from the probability distribution obtained from measuring every basis for the quantum state tomography and reconstruct a sampled state using maximum likelihood estimation. The error of the fidelity is then given by the standard deviation of the fidelity between the sampled states and the ideal GHZ state, while the error of the GME corresponds to the standard deviation of every GME value of the sampled states after the classical optimization with Basin-hopping.

\begin{figure*}[t!]
	\centering
	\begin{subfigure}[b]{0.45\textwidth}
    	\centering
    	\includegraphics[width=\textwidth]{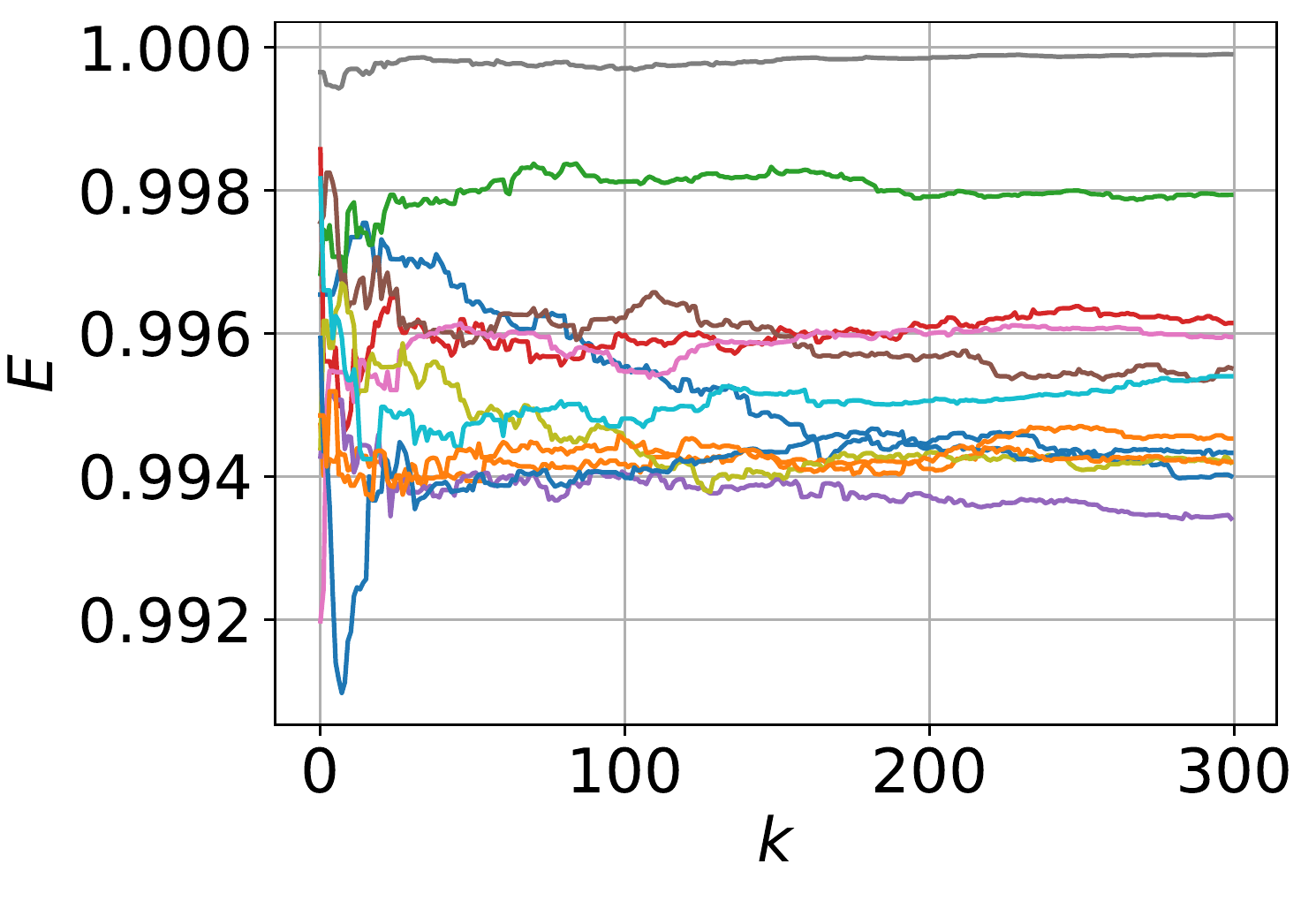}
    	\caption{GME for $|\text{GHZ} \rangle$ using VGDE.}
    	\label{fig:Fig4-1}
    \end{subfigure}
	\begin{subfigure}[b]{0.45\textwidth}
    	\centering
    	\includegraphics[width=\textwidth]{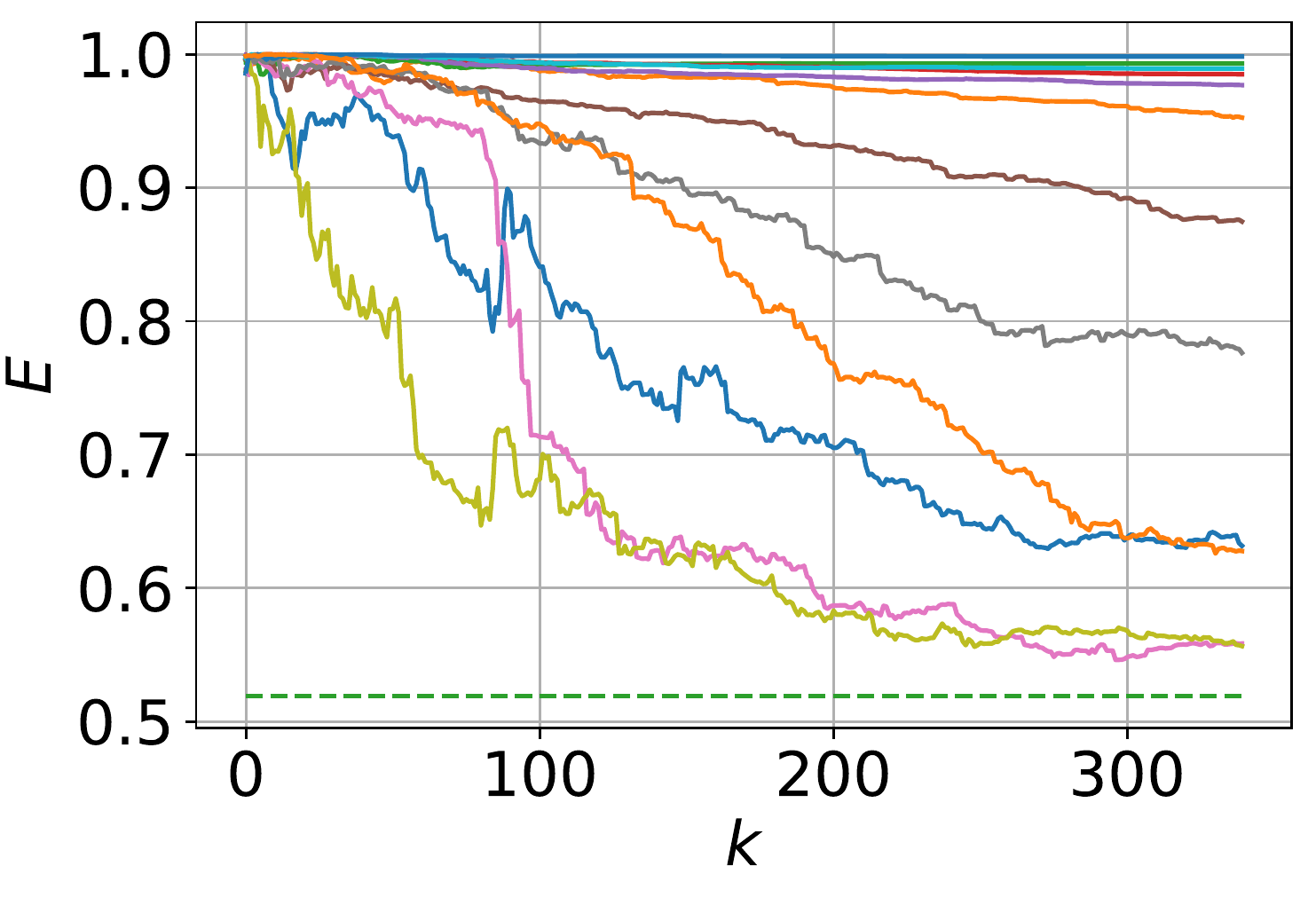}
    	\caption{GME for $|\text{GHZ} \rangle$ using iVDGE.}
    	\label{fig:Fig4-2}
	\end{subfigure}
	\caption{GME value for a 7-qubit $|\text{GHZ} \rangle$ state as a function of the number of iterations $k$ using both VDGE (left) and iVDGE (right) methods. Each solid line corresponds to an optimization routine with different initial conditions. The dotted line corresponds to the exact GME value after performing quantum state tomography on the generated state.}
\label{fig:Fig4}
\end{figure*}

Figure~\ref{fig:Fig4} shows the value of the GME for the experimentally estimated GHZ state provided by (a) VDGE and (b) iVDGE as a function of the number of iterations, for 12 independent realizations. The GME is calculated by evaluating the infidelity between the experimentally estimated GHZ state and the separable state provided by VDGE or iVDGE. The dotted line represents the theoretical GME value of $0.519$. According to Fig.~\ref{fig:Fig4}, of the 12 repetitions of VDGE, not a single one manages to escape the BP, while 5 of the 12 repetitions of iVDGE manage to surpass it, following the criteria defined in our numerical simulations. This percentage of BP appearance is higher than the one expected from the noisy simulations. One possible explanation is the fact that the model does not consider other sources of error, like decoherence, cross-talk between qubits, state preparation error, etc.

\section{\label{sec:6}Discussion}

The geometric measure of entanglement can be evaluated in NISQ devices via the VDGE method. This is done using a parametric quantum circuit which involves local gates only. This allows us to avoid two sources of the BP phenomenon: deep parametric quantum circuits and high levels of noise. However, since the function to be optimized is global, that is, its evaluation requires measurements onto sets of many qubits simultaneously, the BP phenomenon still impairs the performance of VDGE method for a large number of qubits.

In order to avoid the BP phenomenon in the context of the GME we propose the iVDGE method, which adds a previous stage to the VDGE method. This is based on measuring a two-qubit local function, which is the expectation value $\langle H_L \rangle$ of a $2$-qubit Hamiltonian, that upper and lower bounds the global function. This suggests that optimizing the local function allows us to optimize the global one. The local function is minimized for a fixed number of iterations, which produces preliminary parameters that are used thereafter as starting point for VDGE. The first stage of iVDGE can be efficiently carried out using an unbiased estimator $X_g$ with bounded error of the local function. 

We performed several numerical simulations to investigate the performance of iVDGE to estimate the GME and its ability to avoid BP. We start by analyzing both methods for random states in a small number of qubits, as it is shown in Fig.~\ref{fig:Fig1}. Since we are working with a low number of qubits, we can expect both VDGE and iVDGE to converge to the optimum given a suitable number of iterations. Nevertheless, we can see that iVDGE provides a speed-up in the convergence with respect to VDGE for a fixed total number of shots. To assess the performance of the methods for a large number of qubits, we run simulations for superpositions of GHZ and W states of 18 qubits. As Figs.~\ref{fig:Fig2} and \ref{fig:Fig3} indicate, the VDGE method stagnates at $E = 1$, showing the presence of the BP phenomenon. Meanwhile, iVDGE avoids it and successfully converges to the correct value of the GME. Notice that Fig.~\ref{fig:Fig3} shows that the first stage of iVDGE does not necessarily provide an accurate estimate for the GME. Nevertheless, the iVDGE method produces a good enough starting point such that the VDGE method will avoid the flat landscape and converge. Finally, we study the efficiency of iVDGE in presence of noise by simulating a 7-qubit GHZ state using the model of an actual NISQ device, so that a noise-induced BP arises. According to Table~\ref{BP_VDGE}, the VDGE method escapes from the BP with a probability of approximately 0.3, which leads to a median GME value of 0.98, far away from the correct GME value of 0.5, in a wide range of global shot number. The scenario in the case of iVDGE, shown in Table~\ref{BP_iVDGE}, is drastically different, where iVDGE escapes from the BP and converges to the correct GME value with a high probability of approximately 0.8, even for a low global shot number of 256. This shows that iVDGE not only avoids BPs produced by global objective functions but also avoids noise-induced BPs. 

To test the performance of the iVDGE method on the current quantum hardware generation, we conducted experiments on the IBM quantum processor \texttt{ibm\_oslo}. We studied the GME of a 7-qubit GHZ state and to illustrate the difference between the performance of VDGE and iVDGE we choose a low number of shots, such that BP emerges. In particular, we considered a number of 256 shots, which is twice the dimension of the search space. Fig.~\ref{fig:Fig4} shows the estimated GME value of 12 independent runs, where the VDGE method does not escape the BP while iVDGE effectively shows that most of the runs escape the BP and even some of them are close to the correct GME value.


For the experimental results, the dimension of the system is $2^7 = 128$. When we start from a random separable state, the fidelity with respect to the state of the system is $O(1/128)$, and using $256$ shots to estimate this quantity is not enough to construct a good approximation of the gradient. This is one of the effects seen in Fig.~\ref{fig:Fig4}, where none of the trajectories of VDGE left the barren plateau zone. On the other hand, in the first stage of iVDGE we have $128$ shots to estimate fidelities in dimension $2$, which can be done with more precision. Then, when we pass to the second stage, $5$ of $12$ trajectories are able to converge.  

Considering the previous discussion, iVDGE is effectively an improvement of VDGE. We have seen that it is able to avoid the BP phenomena produced by global functions in high dimensional systems, and also the noise-induced BP. Moreover, it provides a speed-up over VDGE in scenarios where the BP phenomenon is not observed. These improvements do not add complexity to the algorithm, as the local function used in the initial stage can be efficiently evaluated and it employs the same ansatz as VDGE. In our simulations, the number of shots required does not scale exponentially with the number of qubits. Then, we expect to have an accurate estimate of the GME even in high-dimensional systems. 

Our protocol finds direct application in estimating the entanglement of many-body quantum states in current devices. For example, recently 27-qubits GHZ states \cite{Mooney_2021} and native-graph states \cite{Mooney_2021_2} have been implemented. The ability of iVDGE to avoid barren plateaus qualifies it as an efficient alternative to certify the entanglement of these states. Moreover, our proposal represents a step forward and complements previous results to handle barren plateaus in variational algorithms. The standard approach is to optimize in the first stage a Hamiltonian composed of single-qubit observables, removing the coupling between the subsystems. In contrast, our approach uses two-qubit observables, so it preserves some of the entanglement structure of the Hamiltonian. This property can be advantageous for problems where entanglement plays a crucial role, such as estimating the ground state of molecules, which are known to be entangled.

The iVDGE method can be enhanced in various ways. The switch between local and global Hamiltonians is crucial in the performance of the algorithm, so establishing a precise criterion for switching without compromising fidelity is essential. This can be done, for instance, by considering a suitable set of gain coefficients on CSPSA. An alternative approach \cite{LaRose2019} has been recently suggested, where an adiabatic process is employed instead of abruptly changing the Hamiltonian from local to global. This is done evaluating the Hamiltonian $H_k=(1-\lambda_k)H_L + \lambda_kH_G$ in the $k$-th iteration of the optimization, where $\{ \lambda_k\}$ is a sequence of positive coefficients such that $\lambda_1=0$ and $\lambda_{k_{\rm max}}=1$. This approach allows us to escape from the flat landscape using the local Hamiltonian in early iterations, while the global Hamiltonian permits precise estimation of GME in later iterations. Furthermore, our protocol is not limited to the use of 2-qubit local Hamiltonians. Adding more stages to the protocol, where local Hamiltonians with more qubits are used, could speed up the method. For example, firstly optimizing over 2-qubits local Hamiltonians, then with $n/2$-qubits local Hamiltonians, and in the last stage carrying out a global optimization with $n$-qubits. These upgrades can enhance the efficiency and accuracy of the protocol, making it a valuable tool for entanglement estimation in many-body quantum systems.





\begin{acknowledgments}
This work was supported by ANID -- Millennium Science Initiative Program -- ICN17$_-$012. L.Z. was supported by the Government of Spain (Severo Ochoa CEX2019-000910-S, TRANQI and European Union NextGenerationEU PRTR-C17.I1), Fundació Cellex, Fundació Mir-Puig and Generalitat de Catalunya (CERCA program). AD was supported by FONDECYT Grants 1231940 and 1230586. MM was supported by ANID-PFCHA/DOCTORADO-NACIONAL/2019-21190958. LP was supported by ANID-PFCHA/DOCTORADO-BECAS-CHILE/2019-772200275, the CSIC Interdisciplinary Thematic Platform (PTI+) on Quantum Technologies (PTI-QTEP+), the CAM/FEDER Project No. S2018/TCS-4342 (QUITEMAD-CM), and the Proyecto Sinérgico CAM 2020 Y2020/TCS-6545 (NanoQuCo-CM).
\end{acknowledgments}

\bibliography{bib}

\begin{thebibliography}{66}%
\makeatletter
\providecommand \@ifxundefined [1]{%
 \@ifx{#1\undefined}
}%
\providecommand \@ifnum [1]{%
 \ifnum #1\expandafter \@firstoftwo
 \else \expandafter \@secondoftwo
 \fi
}%
\providecommand \@ifx [1]{%
 \ifx #1\expandafter \@firstoftwo
 \else \expandafter \@secondoftwo
 \fi
}%
\providecommand \natexlab [1]{#1}%
\providecommand \enquote  [1]{``#1''}%
\providecommand \bibnamefont  [1]{#1}%
\providecommand \bibfnamefont [1]{#1}%
\providecommand \citenamefont [1]{#1}%
\providecommand \href@noop [0]{\@secondoftwo}%
\providecommand \href [0]{\begingroup \@sanitize@url \@href}%
\providecommand \@href[1]{\@@startlink{#1}\@@href}%
\providecommand \@@href[1]{\endgroup#1\@@endlink}%
\providecommand \@sanitize@url [0]{\catcode `\\12\catcode `\$12\catcode
  `\&12\catcode `\#12\catcode `\^12\catcode `\_12\catcode `\%12\relax}%
\providecommand \@@startlink[1]{}%
\providecommand \@@endlink[0]{}%
\providecommand \url  [0]{\begingroup\@sanitize@url \@url }%
\providecommand \@url [1]{\endgroup\@href {#1}{\urlprefix }}%
\providecommand \urlprefix  [0]{URL }%
\providecommand \Eprint [0]{\href }%
\providecommand \doibase [0]{https://doi.org/}%
\providecommand \selectlanguage [0]{\@gobble}%
\providecommand \bibinfo  [0]{\@secondoftwo}%
\providecommand \bibfield  [0]{\@secondoftwo}%
\providecommand \translation [1]{[#1]}%
\providecommand \BibitemOpen [0]{}%
\providecommand \bibitemStop [0]{}%
\providecommand \bibitemNoStop [0]{.\EOS\space}%
\providecommand \EOS [0]{\spacefactor3000\relax}%
\providecommand \BibitemShut  [1]{\csname bibitem#1\endcsname}%
\let\auto@bib@innerbib\@empty
\bibitem [{\citenamefont {Preskill}(2018)}]{Preskill2018quantumcomputingin}%
  \BibitemOpen
  \bibfield  {author} {\bibinfo {author} {\bibfnamefont {J.}~\bibnamefont
  {Preskill}},\ }\bibfield  {title} {\bibinfo {title} {Quantum {C}omputing in
  the {NISQ} era and beyond},\ }\href
  {https://doi.org/10.22331/q-2018-08-06-79} {\bibfield  {journal} {\bibinfo
  {journal} {{Quantum}}\ }\textbf {\bibinfo {volume} {2}},\ \bibinfo {pages}
  {79} (\bibinfo {year} {2018})}\BibitemShut {NoStop}%
\bibitem [{\citenamefont {Krinner}\ \emph {et~al.}(2022)\citenamefont
  {Krinner}, \citenamefont {Lacroix}, \citenamefont {Remm}, \citenamefont
  {Paolo}, \citenamefont {Genois}, \citenamefont {Leroux}, \citenamefont
  {Hellings}, \citenamefont {Lazar}, \citenamefont {Swiadek}, \citenamefont
  {Herrmann}, \citenamefont {Norris}, \citenamefont {Andersen}, \citenamefont
  {M\"{u}ller}, \citenamefont {Blais}, \citenamefont {Eichler},\ and\
  \citenamefont {Wallraff}}]{Krinner2022QEC}%
  \BibitemOpen
  \bibfield  {author} {\bibinfo {author} {\bibfnamefont {S.}~\bibnamefont
  {Krinner}}, \bibinfo {author} {\bibfnamefont {N.}~\bibnamefont {Lacroix}},
  \bibinfo {author} {\bibfnamefont {A.}~\bibnamefont {Remm}}, \bibinfo {author}
  {\bibfnamefont {A.~D.}\ \bibnamefont {Paolo}}, \bibinfo {author}
  {\bibfnamefont {E.}~\bibnamefont {Genois}}, \bibinfo {author} {\bibfnamefont
  {C.}~\bibnamefont {Leroux}}, \bibinfo {author} {\bibfnamefont
  {C.}~\bibnamefont {Hellings}}, \bibinfo {author} {\bibfnamefont
  {S.}~\bibnamefont {Lazar}}, \bibinfo {author} {\bibfnamefont
  {F.}~\bibnamefont {Swiadek}}, \bibinfo {author} {\bibfnamefont
  {J.}~\bibnamefont {Herrmann}}, \bibinfo {author} {\bibfnamefont {G.~J.}\
  \bibnamefont {Norris}}, \bibinfo {author} {\bibfnamefont {C.~K.}\
  \bibnamefont {Andersen}}, \bibinfo {author} {\bibfnamefont {M.}~\bibnamefont
  {M\"{u}ller}}, \bibinfo {author} {\bibfnamefont {A.}~\bibnamefont {Blais}},
  \bibinfo {author} {\bibfnamefont {C.}~\bibnamefont {Eichler}},\ and\ \bibinfo
  {author} {\bibfnamefont {A.}~\bibnamefont {Wallraff}},\ }\bibfield  {title}
  {\bibinfo {title} {Realizing repeated quantum error correction in a
  distance-three surface code},\ }\href
  {https://doi.org/10.1038/s41586-022-04566-8} {\bibfield  {journal} {\bibinfo
  {journal} {Nature}\ }\textbf {\bibinfo {volume} {605}},\ \bibinfo {pages}
  {669} (\bibinfo {year} {2022})}\BibitemShut {NoStop}%
\bibitem [{\citenamefont {Chen}\ \emph {et~al.}(2022)\citenamefont {Chen},
  \citenamefont {Yoder}, \citenamefont {Kim}, \citenamefont {Sundaresan},
  \citenamefont {Srinivasan}, \citenamefont {Li}, \citenamefont {C\'orcoles},
  \citenamefont {Cross},\ and\ \citenamefont {Takita}}]{Chen2022QEC}%
  \BibitemOpen
  \bibfield  {author} {\bibinfo {author} {\bibfnamefont {E.~H.}\ \bibnamefont
  {Chen}}, \bibinfo {author} {\bibfnamefont {T.~J.}\ \bibnamefont {Yoder}},
  \bibinfo {author} {\bibfnamefont {Y.}~\bibnamefont {Kim}}, \bibinfo {author}
  {\bibfnamefont {N.}~\bibnamefont {Sundaresan}}, \bibinfo {author}
  {\bibfnamefont {S.}~\bibnamefont {Srinivasan}}, \bibinfo {author}
  {\bibfnamefont {M.}~\bibnamefont {Li}}, \bibinfo {author} {\bibfnamefont
  {A.~D.}\ \bibnamefont {C\'orcoles}}, \bibinfo {author} {\bibfnamefont
  {A.~W.}\ \bibnamefont {Cross}},\ and\ \bibinfo {author} {\bibfnamefont
  {M.}~\bibnamefont {Takita}},\ }\bibfield  {title} {\bibinfo {title}
  {Calibrated decoders for experimental quantum error correction},\ }\href
  {https://doi.org/10.1103/PhysRevLett.128.110504} {\bibfield  {journal}
  {\bibinfo  {journal} {Phys. Rev. Lett.}\ }\textbf {\bibinfo {volume} {128}},\
  \bibinfo {pages} {110504} (\bibinfo {year} {2022})}\BibitemShut {NoStop}%
\bibitem [{\citenamefont {Cerezo}\ \emph
  {et~al.}(2021{\natexlab{a}})\citenamefont {Cerezo}, \citenamefont
  {Arrasmith}, \citenamefont {Babbush}, \citenamefont {Benjamin}, \citenamefont
  {Endo}, \citenamefont {Fujii}, \citenamefont {McClean}, \citenamefont
  {Mitarai}, \citenamefont {Yuan}, \citenamefont {Cincio},\ and\ \citenamefont
  {Coles}}]{Cerezo2021VQA}%
  \BibitemOpen
  \bibfield  {author} {\bibinfo {author} {\bibfnamefont {M.}~\bibnamefont
  {Cerezo}}, \bibinfo {author} {\bibfnamefont {A.}~\bibnamefont {Arrasmith}},
  \bibinfo {author} {\bibfnamefont {R.}~\bibnamefont {Babbush}}, \bibinfo
  {author} {\bibfnamefont {S.~C.}\ \bibnamefont {Benjamin}}, \bibinfo {author}
  {\bibfnamefont {S.}~\bibnamefont {Endo}}, \bibinfo {author} {\bibfnamefont
  {K.}~\bibnamefont {Fujii}}, \bibinfo {author} {\bibfnamefont {J.~R.}\
  \bibnamefont {McClean}}, \bibinfo {author} {\bibfnamefont {K.}~\bibnamefont
  {Mitarai}}, \bibinfo {author} {\bibfnamefont {X.}~\bibnamefont {Yuan}},
  \bibinfo {author} {\bibfnamefont {L.}~\bibnamefont {Cincio}},\ and\ \bibinfo
  {author} {\bibfnamefont {P.~J.}\ \bibnamefont {Coles}},\ }\bibfield  {title}
  {\bibinfo {title} {Variational quantum algorithms},\ }\href
  {https://doi.org/10.1038/s42254-021-00348-9} {\bibfield  {journal} {\bibinfo
  {journal} {Nature Reviews Physics}\ }\textbf {\bibinfo {volume} {3}},\
  \bibinfo {pages} {625} (\bibinfo {year} {2021}{\natexlab{a}})}\BibitemShut
  {NoStop}%
\bibitem [{\citenamefont {Bharti}\ \emph {et~al.}(2022)\citenamefont {Bharti},
  \citenamefont {Cervera-Lierta}, \citenamefont {Kyaw}, \citenamefont {Haug},
  \citenamefont {Alperin-Lea}, \citenamefont {Anand}, \citenamefont {Degroote},
  \citenamefont {Heimonen}, \citenamefont {Kottmann}, \citenamefont {Menke},
  \citenamefont {Mok}, \citenamefont {Sim}, \citenamefont {Kwek},\ and\
  \citenamefont {Aspuru-Guzik}}]{Bharti2022VQA}%
  \BibitemOpen
  \bibfield  {author} {\bibinfo {author} {\bibfnamefont {K.}~\bibnamefont
  {Bharti}}, \bibinfo {author} {\bibfnamefont {A.}~\bibnamefont
  {Cervera-Lierta}}, \bibinfo {author} {\bibfnamefont {T.~H.}\ \bibnamefont
  {Kyaw}}, \bibinfo {author} {\bibfnamefont {T.}~\bibnamefont {Haug}}, \bibinfo
  {author} {\bibfnamefont {S.}~\bibnamefont {Alperin-Lea}}, \bibinfo {author}
  {\bibfnamefont {A.}~\bibnamefont {Anand}}, \bibinfo {author} {\bibfnamefont
  {M.}~\bibnamefont {Degroote}}, \bibinfo {author} {\bibfnamefont
  {H.}~\bibnamefont {Heimonen}}, \bibinfo {author} {\bibfnamefont {J.~S.}\
  \bibnamefont {Kottmann}}, \bibinfo {author} {\bibfnamefont {T.}~\bibnamefont
  {Menke}}, \bibinfo {author} {\bibfnamefont {W.-K.}\ \bibnamefont {Mok}},
  \bibinfo {author} {\bibfnamefont {S.}~\bibnamefont {Sim}}, \bibinfo {author}
  {\bibfnamefont {L.-C.}\ \bibnamefont {Kwek}},\ and\ \bibinfo {author}
  {\bibfnamefont {A.}~\bibnamefont {Aspuru-Guzik}},\ }\bibfield  {title}
  {\bibinfo {title} {Noisy intermediate-scale quantum algorithms},\ }\href
  {https://doi.org/10.1103/RevModPhys.94.015004} {\bibfield  {journal}
  {\bibinfo  {journal} {Rev. Mod. Phys.}\ }\textbf {\bibinfo {volume} {94}},\
  \bibinfo {pages} {015004} (\bibinfo {year} {2022})}\BibitemShut {NoStop}%
\bibitem [{\citenamefont {Tilly}\ \emph {et~al.}(2022)\citenamefont {Tilly},
  \citenamefont {Chen}, \citenamefont {Cao}, \citenamefont {Picozzi},
  \citenamefont {Setia}, \citenamefont {Li}, \citenamefont {Grant},
  \citenamefont {Wossnig}, \citenamefont {Rungger}, \citenamefont {Booth},\
  and\ \citenamefont {Tennyson}}]{Tilly2022VQA}%
  \BibitemOpen
  \bibfield  {author} {\bibinfo {author} {\bibfnamefont {J.}~\bibnamefont
  {Tilly}}, \bibinfo {author} {\bibfnamefont {H.}~\bibnamefont {Chen}},
  \bibinfo {author} {\bibfnamefont {S.}~\bibnamefont {Cao}}, \bibinfo {author}
  {\bibfnamefont {D.}~\bibnamefont {Picozzi}}, \bibinfo {author} {\bibfnamefont
  {K.}~\bibnamefont {Setia}}, \bibinfo {author} {\bibfnamefont
  {Y.}~\bibnamefont {Li}}, \bibinfo {author} {\bibfnamefont {E.}~\bibnamefont
  {Grant}}, \bibinfo {author} {\bibfnamefont {L.}~\bibnamefont {Wossnig}},
  \bibinfo {author} {\bibfnamefont {I.}~\bibnamefont {Rungger}}, \bibinfo
  {author} {\bibfnamefont {G.~H.}\ \bibnamefont {Booth}},\ and\ \bibinfo
  {author} {\bibfnamefont {J.}~\bibnamefont {Tennyson}},\ }\bibfield  {title}
  {\bibinfo {title} {The variational quantum eigensolver: A review of methods
  and best practices},\ }\href {https://doi.org/10.1016/j.physrep.2022.08.003}
  {\bibfield  {journal} {\bibinfo  {journal} {Physics Reports}\ }\textbf
  {\bibinfo {volume} {986}},\ \bibinfo {pages} {1} (\bibinfo {year}
  {2022})}\BibitemShut {NoStop}%
\bibitem [{\citenamefont {Peruzzo}\ \emph {et~al.}(2014)\citenamefont
  {Peruzzo}, \citenamefont {McClean}, \citenamefont {Shadbolt}, \citenamefont
  {Yung}, \citenamefont {Zhou}, \citenamefont {Love}, \citenamefont
  {Aspuru-Guzik},\ and\ \citenamefont {O'Brien}}]{Peruzzo2014VQE}%
  \BibitemOpen
  \bibfield  {author} {\bibinfo {author} {\bibfnamefont {A.}~\bibnamefont
  {Peruzzo}}, \bibinfo {author} {\bibfnamefont {J.}~\bibnamefont {McClean}},
  \bibinfo {author} {\bibfnamefont {P.}~\bibnamefont {Shadbolt}}, \bibinfo
  {author} {\bibfnamefont {M.-H.}\ \bibnamefont {Yung}}, \bibinfo {author}
  {\bibfnamefont {X.-Q.}\ \bibnamefont {Zhou}}, \bibinfo {author}
  {\bibfnamefont {P.~J.}\ \bibnamefont {Love}}, \bibinfo {author}
  {\bibfnamefont {A.}~\bibnamefont {Aspuru-Guzik}},\ and\ \bibinfo {author}
  {\bibfnamefont {J.~L.}\ \bibnamefont {O'Brien}},\ }\bibfield  {title}
  {\bibinfo {title} {A variational eigenvalue solver on a photonic quantum
  processor},\ }\href {https://doi.org/10.1038/ncomms5213} {\bibfield
  {journal} {\bibinfo  {journal} {Nature Communications}\ }\textbf {\bibinfo
  {volume} {5}},\ \bibinfo {pages} {4213} (\bibinfo {year} {2014})}\BibitemShut
  {NoStop}%
\bibitem [{\citenamefont {Farhi}\ \emph {et~al.}(2009)\citenamefont {Farhi},
  \citenamefont {Goldstone},\ and\ \citenamefont {Gutmann}}]{Farhi2014QAOA}%
  \BibitemOpen
  \bibfield  {author} {\bibinfo {author} {\bibfnamefont {E.}~\bibnamefont
  {Farhi}}, \bibinfo {author} {\bibfnamefont {J.}~\bibnamefont {Goldstone}},\
  and\ \bibinfo {author} {\bibfnamefont {S.}~\bibnamefont {Gutmann}},\ }\href
  {https://doi.org/10.48550/ARXIV.1411.4028} {\bibinfo {title} {A quantum
  approximate optimization algorithm}} (\bibinfo {year} {2009}),\ \Eprint
  {https://arxiv.org/abs/1411.4028} {arXiv:1411.4028} \BibitemShut {NoStop}%
\bibitem [{\citenamefont {Biamonte}\ \emph {et~al.}(2017)\citenamefont
  {Biamonte}, \citenamefont {Wittek}, \citenamefont {Pancotti}, \citenamefont
  {Rebentrost}, \citenamefont {Wiebe},\ and\ \citenamefont
  {Lloyd}}]{Biamonte2017}%
  \BibitemOpen
  \bibfield  {author} {\bibinfo {author} {\bibfnamefont {J.}~\bibnamefont
  {Biamonte}}, \bibinfo {author} {\bibfnamefont {P.}~\bibnamefont {Wittek}},
  \bibinfo {author} {\bibfnamefont {N.}~\bibnamefont {Pancotti}}, \bibinfo
  {author} {\bibfnamefont {P.}~\bibnamefont {Rebentrost}}, \bibinfo {author}
  {\bibfnamefont {N.}~\bibnamefont {Wiebe}},\ and\ \bibinfo {author}
  {\bibfnamefont {S.}~\bibnamefont {Lloyd}},\ }\bibfield  {title} {\bibinfo
  {title} {Quantum machine learning},\ }\href
  {https://doi.org/10.1038/nature23474} {\bibfield  {journal} {\bibinfo
  {journal} {Nature}\ }\textbf {\bibinfo {volume} {549}},\ \bibinfo {pages}
  {195} (\bibinfo {year} {2017})}\BibitemShut {NoStop}%
\bibitem [{\citenamefont {Beer}\ \emph {et~al.}(2020)\citenamefont {Beer},
  \citenamefont {Bondarenko}, \citenamefont {Farrelly}, \citenamefont
  {Osborne}, \citenamefont {Salzmann}, \citenamefont {Scheiermann},\ and\
  \citenamefont {Wolf}}]{Beer2020}%
  \BibitemOpen
  \bibfield  {author} {\bibinfo {author} {\bibfnamefont {K.}~\bibnamefont
  {Beer}}, \bibinfo {author} {\bibfnamefont {D.}~\bibnamefont {Bondarenko}},
  \bibinfo {author} {\bibfnamefont {T.}~\bibnamefont {Farrelly}}, \bibinfo
  {author} {\bibfnamefont {T.~J.}\ \bibnamefont {Osborne}}, \bibinfo {author}
  {\bibfnamefont {R.}~\bibnamefont {Salzmann}}, \bibinfo {author}
  {\bibfnamefont {D.}~\bibnamefont {Scheiermann}},\ and\ \bibinfo {author}
  {\bibfnamefont {R.}~\bibnamefont {Wolf}},\ }\bibfield  {title} {\bibinfo
  {title} {Training deep quantum neural networks},\ }\href
  {https://doi.org/10.1038/s41467-020-14454-2} {\bibfield  {journal} {\bibinfo
  {journal} {Nature Communications}\ }\textbf {\bibinfo {volume} {11}},\
  \bibinfo {pages} {808} (\bibinfo {year} {2020})}\BibitemShut {NoStop}%
\bibitem [{\citenamefont {Abbas}\ \emph {et~al.}(2021)\citenamefont {Abbas},
  \citenamefont {Sutter}, \citenamefont {Zoufal}, \citenamefont {Lucchi},
  \citenamefont {Figalli},\ and\ \citenamefont {Woerner}}]{Abbas2021}%
  \BibitemOpen
  \bibfield  {author} {\bibinfo {author} {\bibfnamefont {A.}~\bibnamefont
  {Abbas}}, \bibinfo {author} {\bibfnamefont {D.}~\bibnamefont {Sutter}},
  \bibinfo {author} {\bibfnamefont {C.}~\bibnamefont {Zoufal}}, \bibinfo
  {author} {\bibfnamefont {A.}~\bibnamefont {Lucchi}}, \bibinfo {author}
  {\bibfnamefont {A.}~\bibnamefont {Figalli}},\ and\ \bibinfo {author}
  {\bibfnamefont {S.}~\bibnamefont {Woerner}},\ }\bibfield  {title} {\bibinfo
  {title} {The power of quantum neural networks},\ }\href
  {https://doi.org/10.1038/s43588-021-00084-1} {\bibfield  {journal} {\bibinfo
  {journal} {Nature Computational Science}\ }\textbf {\bibinfo {volume} {1}},\
  \bibinfo {pages} {403} (\bibinfo {year} {2021})}\BibitemShut {NoStop}%
\bibitem [{\citenamefont {Lanyon}\ \emph {et~al.}(2010)\citenamefont {Lanyon},
  \citenamefont {Whitfield}, \citenamefont {Gillett}, \citenamefont {Goggin},
  \citenamefont {Almeida}, \citenamefont {Kassal}, \citenamefont {Biamonte},
  \citenamefont {Mohseni}, \citenamefont {Powell}, \citenamefont {Barbieri},
  \citenamefont {Aspuru-Guzik},\ and\ \citenamefont {White}}]{Lanyon2010chem}%
  \BibitemOpen
  \bibfield  {author} {\bibinfo {author} {\bibfnamefont {B.~P.}\ \bibnamefont
  {Lanyon}}, \bibinfo {author} {\bibfnamefont {J.~D.}\ \bibnamefont
  {Whitfield}}, \bibinfo {author} {\bibfnamefont {G.~G.}\ \bibnamefont
  {Gillett}}, \bibinfo {author} {\bibfnamefont {M.~E.}\ \bibnamefont {Goggin}},
  \bibinfo {author} {\bibfnamefont {M.~P.}\ \bibnamefont {Almeida}}, \bibinfo
  {author} {\bibfnamefont {I.}~\bibnamefont {Kassal}}, \bibinfo {author}
  {\bibfnamefont {J.~D.}\ \bibnamefont {Biamonte}}, \bibinfo {author}
  {\bibfnamefont {M.}~\bibnamefont {Mohseni}}, \bibinfo {author} {\bibfnamefont
  {B.~J.}\ \bibnamefont {Powell}}, \bibinfo {author} {\bibfnamefont
  {M.}~\bibnamefont {Barbieri}}, \bibinfo {author} {\bibfnamefont
  {A.}~\bibnamefont {Aspuru-Guzik}},\ and\ \bibinfo {author} {\bibfnamefont
  {A.~G.}\ \bibnamefont {White}},\ }\bibfield  {title} {\bibinfo {title}
  {Towards quantum chemistry on a quantum computer},\ }\href
  {https://doi.org/10.1038/nchem.483} {\bibfield  {journal} {\bibinfo
  {journal} {Nature Chemistry}\ }\textbf {\bibinfo {volume} {2}},\ \bibinfo
  {pages} {106} (\bibinfo {year} {2010})}\BibitemShut {NoStop}%
\bibitem [{\citenamefont {Hempel}\ \emph {et~al.}(2018)\citenamefont {Hempel},
  \citenamefont {Maier}, \citenamefont {Romero}, \citenamefont {McClean},
  \citenamefont {Monz}, \citenamefont {Shen}, \citenamefont {Jurcevic},
  \citenamefont {Lanyon}, \citenamefont {Love}, \citenamefont {Babbush},
  \citenamefont {Aspuru-Guzik}, \citenamefont {Blatt},\ and\ \citenamefont
  {Roos}}]{Hempel2018chem}%
  \BibitemOpen
  \bibfield  {author} {\bibinfo {author} {\bibfnamefont {C.}~\bibnamefont
  {Hempel}}, \bibinfo {author} {\bibfnamefont {C.}~\bibnamefont {Maier}},
  \bibinfo {author} {\bibfnamefont {J.}~\bibnamefont {Romero}}, \bibinfo
  {author} {\bibfnamefont {J.}~\bibnamefont {McClean}}, \bibinfo {author}
  {\bibfnamefont {T.}~\bibnamefont {Monz}}, \bibinfo {author} {\bibfnamefont
  {H.}~\bibnamefont {Shen}}, \bibinfo {author} {\bibfnamefont {P.}~\bibnamefont
  {Jurcevic}}, \bibinfo {author} {\bibfnamefont {B.~P.}\ \bibnamefont
  {Lanyon}}, \bibinfo {author} {\bibfnamefont {P.}~\bibnamefont {Love}},
  \bibinfo {author} {\bibfnamefont {R.}~\bibnamefont {Babbush}}, \bibinfo
  {author} {\bibfnamefont {A.}~\bibnamefont {Aspuru-Guzik}}, \bibinfo {author}
  {\bibfnamefont {R.}~\bibnamefont {Blatt}},\ and\ \bibinfo {author}
  {\bibfnamefont {C.~F.}\ \bibnamefont {Roos}},\ }\bibfield  {title} {\bibinfo
  {title} {Quantum chemistry calculations on a trapped-ion quantum simulator},\
  }\href {https://doi.org/10.1103/PhysRevX.8.031022} {\bibfield  {journal}
  {\bibinfo  {journal} {Phys. Rev. X}\ }\textbf {\bibinfo {volume} {8}},\
  \bibinfo {pages} {031022} (\bibinfo {year} {2018})}\BibitemShut {NoStop}%
\bibitem [{\citenamefont {Nam}\ \emph {et~al.}(2020)\citenamefont {Nam},
  \citenamefont {Chen}, \citenamefont {Pisenti}, \citenamefont {Wright},
  \citenamefont {Delaney}, \citenamefont {Maslov}, \citenamefont {Brown},
  \citenamefont {Allen}, \citenamefont {Amini}, \citenamefont {Apisdorf},
  \citenamefont {Beck}, \citenamefont {Blinov}, \citenamefont {Chaplin},
  \citenamefont {Chmielewski}, \citenamefont {Collins}, \citenamefont
  {Debnath}, \citenamefont {Hudek}, \citenamefont {Ducore}, \citenamefont
  {Keesan}, \citenamefont {Kreikemeier}, \citenamefont {Mizrahi}, \citenamefont
  {Solomon}, \citenamefont {Williams}, \citenamefont {Wong-Campos},
  \citenamefont {Moehring}, \citenamefont {Monroe},\ and\ \citenamefont
  {Kim}}]{Nam2020chem}%
  \BibitemOpen
  \bibfield  {author} {\bibinfo {author} {\bibfnamefont {Y.}~\bibnamefont
  {Nam}}, \bibinfo {author} {\bibfnamefont {J.-S.}\ \bibnamefont {Chen}},
  \bibinfo {author} {\bibfnamefont {N.~C.}\ \bibnamefont {Pisenti}}, \bibinfo
  {author} {\bibfnamefont {K.}~\bibnamefont {Wright}}, \bibinfo {author}
  {\bibfnamefont {C.}~\bibnamefont {Delaney}}, \bibinfo {author} {\bibfnamefont
  {D.}~\bibnamefont {Maslov}}, \bibinfo {author} {\bibfnamefont {K.~R.}\
  \bibnamefont {Brown}}, \bibinfo {author} {\bibfnamefont {S.}~\bibnamefont
  {Allen}}, \bibinfo {author} {\bibfnamefont {J.~M.}\ \bibnamefont {Amini}},
  \bibinfo {author} {\bibfnamefont {J.}~\bibnamefont {Apisdorf}}, \bibinfo
  {author} {\bibfnamefont {K.~M.}\ \bibnamefont {Beck}}, \bibinfo {author}
  {\bibfnamefont {A.}~\bibnamefont {Blinov}}, \bibinfo {author} {\bibfnamefont
  {V.}~\bibnamefont {Chaplin}}, \bibinfo {author} {\bibfnamefont
  {M.}~\bibnamefont {Chmielewski}}, \bibinfo {author} {\bibfnamefont
  {C.}~\bibnamefont {Collins}}, \bibinfo {author} {\bibfnamefont
  {S.}~\bibnamefont {Debnath}}, \bibinfo {author} {\bibfnamefont {K.~M.}\
  \bibnamefont {Hudek}}, \bibinfo {author} {\bibfnamefont {A.~M.}\ \bibnamefont
  {Ducore}}, \bibinfo {author} {\bibfnamefont {M.}~\bibnamefont {Keesan}},
  \bibinfo {author} {\bibfnamefont {S.~M.}\ \bibnamefont {Kreikemeier}},
  \bibinfo {author} {\bibfnamefont {J.}~\bibnamefont {Mizrahi}}, \bibinfo
  {author} {\bibfnamefont {P.}~\bibnamefont {Solomon}}, \bibinfo {author}
  {\bibfnamefont {M.}~\bibnamefont {Williams}}, \bibinfo {author}
  {\bibfnamefont {J.~D.}\ \bibnamefont {Wong-Campos}}, \bibinfo {author}
  {\bibfnamefont {D.}~\bibnamefont {Moehring}}, \bibinfo {author}
  {\bibfnamefont {C.}~\bibnamefont {Monroe}},\ and\ \bibinfo {author}
  {\bibfnamefont {J.}~\bibnamefont {Kim}},\ }\bibfield  {title} {\bibinfo
  {title} {Ground-state energy estimation of the water molecule on a
  trapped-ion quantum computer},\ }\href
  {https://doi.org/10.1038/s41534-020-0259-3} {\bibfield  {journal} {\bibinfo
  {journal} {npj Quantum Information}\ }\textbf {\bibinfo {volume} {6}},\
  \bibinfo {pages} {33} (\bibinfo {year} {2020})}\BibitemShut {NoStop}%
\bibitem [{\citenamefont {Hong}\ \emph {et~al.}(2014)\citenamefont {Hong},
  \citenamefont {Hu},\ and\ \citenamefont {Liu}}]{Hong2014}%
  \BibitemOpen
  \bibfield  {author} {\bibinfo {author} {\bibfnamefont {L.~J.}\ \bibnamefont
  {Hong}}, \bibinfo {author} {\bibfnamefont {Z.}~\bibnamefont {Hu}},\ and\
  \bibinfo {author} {\bibfnamefont {G.}~\bibnamefont {Liu}},\ }\bibfield
  {title} {\bibinfo {title} {Monte carlo methods for value-at-risk and
  conditional value-at-risk},\ }\href {https://doi.org/10.1145/2661631}
  {\bibfield  {journal} {\bibinfo  {journal} {{ACM} Transactions on Modeling
  and Computer Simulation}\ }\textbf {\bibinfo {volume} {24}},\ \bibinfo
  {pages} {1} (\bibinfo {year} {2014})}\BibitemShut {NoStop}%
\bibitem [{\citenamefont {Egger}\ \emph {et~al.}(2020)\citenamefont {Egger},
  \citenamefont {Gambella}, \citenamefont {Marecek}, \citenamefont {McFaddin},
  \citenamefont {Mevissen}, \citenamefont {Raymond}, \citenamefont {Simonetto},
  \citenamefont {Woerner},\ and\ \citenamefont {Yndurain}}]{Egger2020}%
  \BibitemOpen
  \bibfield  {author} {\bibinfo {author} {\bibfnamefont {D.~J.}\ \bibnamefont
  {Egger}}, \bibinfo {author} {\bibfnamefont {C.}~\bibnamefont {Gambella}},
  \bibinfo {author} {\bibfnamefont {J.}~\bibnamefont {Marecek}}, \bibinfo
  {author} {\bibfnamefont {S.}~\bibnamefont {McFaddin}}, \bibinfo {author}
  {\bibfnamefont {M.}~\bibnamefont {Mevissen}}, \bibinfo {author}
  {\bibfnamefont {R.}~\bibnamefont {Raymond}}, \bibinfo {author} {\bibfnamefont
  {A.}~\bibnamefont {Simonetto}}, \bibinfo {author} {\bibfnamefont
  {S.}~\bibnamefont {Woerner}},\ and\ \bibinfo {author} {\bibfnamefont
  {E.}~\bibnamefont {Yndurain}},\ }\bibfield  {title} {\bibinfo {title}
  {Quantum computing for finance: State-of-the-art and future prospects},\
  }\href {https://doi.org/10.1109/TQE.2020.3030314} {\bibfield  {journal}
  {\bibinfo  {journal} {IEEE Transactions on Quantum Engineering}\ }\textbf
  {\bibinfo {volume} {1}},\ \bibinfo {pages} {1} (\bibinfo {year}
  {2020})}\BibitemShut {NoStop}%
\bibitem [{\citenamefont {Barkoutsos}\ \emph {et~al.}(2020)\citenamefont
  {Barkoutsos}, \citenamefont {Nannicini}, \citenamefont {Robert},
  \citenamefont {Tavernelli},\ and\ \citenamefont {Woerner}}]{Barkoutsos2020}%
  \BibitemOpen
  \bibfield  {author} {\bibinfo {author} {\bibfnamefont {P.~K.}\ \bibnamefont
  {Barkoutsos}}, \bibinfo {author} {\bibfnamefont {G.}~\bibnamefont
  {Nannicini}}, \bibinfo {author} {\bibfnamefont {A.}~\bibnamefont {Robert}},
  \bibinfo {author} {\bibfnamefont {I.}~\bibnamefont {Tavernelli}},\ and\
  \bibinfo {author} {\bibfnamefont {S.}~\bibnamefont {Woerner}},\ }\bibfield
  {title} {\bibinfo {title} {Improving variational quantum optimization using
  {CVaR}},\ }\href {https://doi.org/10.22331/q-2020-04-20-256} {\bibfield
  {journal} {\bibinfo  {journal} {Quantum}\ }\textbf {\bibinfo {volume} {4}},\
  \bibinfo {pages} {256} (\bibinfo {year} {2020})}\BibitemShut {NoStop}%
\bibitem [{\citenamefont {Herman}\ \emph {et~al.}(2022)\citenamefont {Herman},
  \citenamefont {Googin}, \citenamefont {Liu}, \citenamefont {Galda},
  \citenamefont {Safro}, \citenamefont {Sun}, \citenamefont {Pistoia},\ and\
  \citenamefont {Alexeev}}]{Herman2022}%
  \BibitemOpen
  \bibfield  {author} {\bibinfo {author} {\bibfnamefont {D.}~\bibnamefont
  {Herman}}, \bibinfo {author} {\bibfnamefont {C.}~\bibnamefont {Googin}},
  \bibinfo {author} {\bibfnamefont {X.}~\bibnamefont {Liu}}, \bibinfo {author}
  {\bibfnamefont {A.}~\bibnamefont {Galda}}, \bibinfo {author} {\bibfnamefont
  {I.}~\bibnamefont {Safro}}, \bibinfo {author} {\bibfnamefont
  {Y.}~\bibnamefont {Sun}}, \bibinfo {author} {\bibfnamefont {M.}~\bibnamefont
  {Pistoia}},\ and\ \bibinfo {author} {\bibfnamefont {Y.}~\bibnamefont
  {Alexeev}},\ }\href {https://doi.org/10.48550/ARXIV.2201.02773} {\bibinfo
  {title} {A survey of quantum computing for finance}} (\bibinfo {year}
  {2022})\BibitemShut {NoStop}%
\bibitem [{\citenamefont {Ferrie}(2014)}]{Ferrie2014}%
  \BibitemOpen
  \bibfield  {author} {\bibinfo {author} {\bibfnamefont {C.}~\bibnamefont
  {Ferrie}},\ }\bibfield  {title} {\bibinfo {title} {Self-guided quantum
  tomography},\ }\href {https://doi.org/10.1103/PhysRevLett.113.190404}
  {\bibfield  {journal} {\bibinfo  {journal} {Phys. Rev. Lett.}\ }\textbf
  {\bibinfo {volume} {113}},\ \bibinfo {pages} {190404} (\bibinfo {year}
  {2014})}\BibitemShut {NoStop}%
\bibitem [{\citenamefont {Chapman}\ \emph {et~al.}(2016)\citenamefont
  {Chapman}, \citenamefont {Ferrie},\ and\ \citenamefont
  {Peruzzo}}]{Chapman2016}%
  \BibitemOpen
  \bibfield  {author} {\bibinfo {author} {\bibfnamefont {R.~J.}\ \bibnamefont
  {Chapman}}, \bibinfo {author} {\bibfnamefont {C.}~\bibnamefont {Ferrie}},\
  and\ \bibinfo {author} {\bibfnamefont {A.}~\bibnamefont {Peruzzo}},\
  }\bibfield  {title} {\bibinfo {title} {Experimental demonstration of
  self-guided quantum tomography},\ }\href
  {https://doi.org/10.1103/PhysRevLett.117.040402} {\bibfield  {journal}
  {\bibinfo  {journal} {Phys. Rev. Lett.}\ }\textbf {\bibinfo {volume} {117}},\
  \bibinfo {pages} {040402} (\bibinfo {year} {2016})}\BibitemShut {NoStop}%
\bibitem [{\citenamefont {Utreras-Alarc{\'{o}}n}\ \emph
  {et~al.}(2019)\citenamefont {Utreras-Alarc{\'{o}}n}, \citenamefont
  {Rivera-Tapia}, \citenamefont {Niklitschek},\ and\ \citenamefont
  {Delgado}}]{UtrerasAlarcn2019}%
  \BibitemOpen
  \bibfield  {author} {\bibinfo {author} {\bibfnamefont {A.}~\bibnamefont
  {Utreras-Alarc{\'{o}}n}}, \bibinfo {author} {\bibfnamefont {M.}~\bibnamefont
  {Rivera-Tapia}}, \bibinfo {author} {\bibfnamefont {S.}~\bibnamefont
  {Niklitschek}},\ and\ \bibinfo {author} {\bibfnamefont {A.}~\bibnamefont
  {Delgado}},\ }\bibfield  {title} {\bibinfo {title} {Stochastic optimization
  on complex variables and pure-state quantum tomography},\ }\href
  {https://doi.org/10.1038/s41598-019-52289-0} {\bibfield  {journal} {\bibinfo
  {journal} {Scientific Reports}\ }\textbf {\bibinfo {volume} {9}},\ \bibinfo
  {pages} {16143} (\bibinfo {year} {2019})}\BibitemShut {NoStop}%
\bibitem [{\citenamefont {Zambrano}\ \emph {et~al.}(2020)\citenamefont
  {Zambrano}, \citenamefont {Pereira}, \citenamefont {Niklitschek},\ and\
  \citenamefont {Delgado}}]{Zambrano2020}%
  \BibitemOpen
  \bibfield  {author} {\bibinfo {author} {\bibfnamefont {L.}~\bibnamefont
  {Zambrano}}, \bibinfo {author} {\bibfnamefont {L.}~\bibnamefont {Pereira}},
  \bibinfo {author} {\bibfnamefont {S.}~\bibnamefont {Niklitschek}},\ and\
  \bibinfo {author} {\bibfnamefont {A.}~\bibnamefont {Delgado}},\ }\bibfield
  {title} {\bibinfo {title} {Estimation of pure quantum states in high
  dimension at the limit of quantum accuracy through complex optimization and
  statistical inference},\ }\href {https://doi.org/10.1038/s41598-020-69646-z}
  {\bibfield  {journal} {\bibinfo  {journal} {Scientific Reports}\ }\textbf
  {\bibinfo {volume} {10}},\ \bibinfo {pages} {12781} (\bibinfo {year}
  {2020})}\BibitemShut {NoStop}%
\bibitem [{\citenamefont {Liu}\ \emph {et~al.}(2020)\citenamefont {Liu},
  \citenamefont {Wang}, \citenamefont {Xue}, \citenamefont {Huang},
  \citenamefont {Fu}, \citenamefont {Qiang}, \citenamefont {Xu}, \citenamefont
  {Huang}, \citenamefont {Deng}, \citenamefont {Guo}, \citenamefont {Yang},\
  and\ \citenamefont {Wu}}]{Liu2020}%
  \BibitemOpen
  \bibfield  {author} {\bibinfo {author} {\bibfnamefont {Y.}~\bibnamefont
  {Liu}}, \bibinfo {author} {\bibfnamefont {D.}~\bibnamefont {Wang}}, \bibinfo
  {author} {\bibfnamefont {S.}~\bibnamefont {Xue}}, \bibinfo {author}
  {\bibfnamefont {A.}~\bibnamefont {Huang}}, \bibinfo {author} {\bibfnamefont
  {X.}~\bibnamefont {Fu}}, \bibinfo {author} {\bibfnamefont {X.}~\bibnamefont
  {Qiang}}, \bibinfo {author} {\bibfnamefont {P.}~\bibnamefont {Xu}}, \bibinfo
  {author} {\bibfnamefont {H.-L.}\ \bibnamefont {Huang}}, \bibinfo {author}
  {\bibfnamefont {M.}~\bibnamefont {Deng}}, \bibinfo {author} {\bibfnamefont
  {C.}~\bibnamefont {Guo}}, \bibinfo {author} {\bibfnamefont {X.}~\bibnamefont
  {Yang}},\ and\ \bibinfo {author} {\bibfnamefont {J.}~\bibnamefont {Wu}},\
  }\bibfield  {title} {\bibinfo {title} {Variational quantum circuits for
  quantum state tomography},\ }\href
  {https://doi.org/10.1103/PhysRevA.101.052316} {\bibfield  {journal} {\bibinfo
   {journal} {Phys. Rev. A}\ }\textbf {\bibinfo {volume} {101}},\ \bibinfo
  {pages} {052316} (\bibinfo {year} {2020})}\BibitemShut {NoStop}%
\bibitem [{\citenamefont {Rambach}\ \emph {et~al.}(2021)\citenamefont
  {Rambach}, \citenamefont {Qaryan}, \citenamefont {Kewming}, \citenamefont
  {Ferrie}, \citenamefont {White},\ and\ \citenamefont {Romero}}]{Rambach2021}%
  \BibitemOpen
  \bibfield  {author} {\bibinfo {author} {\bibfnamefont {M.}~\bibnamefont
  {Rambach}}, \bibinfo {author} {\bibfnamefont {M.}~\bibnamefont {Qaryan}},
  \bibinfo {author} {\bibfnamefont {M.}~\bibnamefont {Kewming}}, \bibinfo
  {author} {\bibfnamefont {C.}~\bibnamefont {Ferrie}}, \bibinfo {author}
  {\bibfnamefont {A.~G.}\ \bibnamefont {White}},\ and\ \bibinfo {author}
  {\bibfnamefont {J.}~\bibnamefont {Romero}},\ }\bibfield  {title} {\bibinfo
  {title} {Robust and efficient high-dimensional quantum state tomography},\
  }\href {https://doi.org/10.1103/PhysRevLett.126.100402} {\bibfield  {journal}
  {\bibinfo  {journal} {Phys. Rev. Lett.}\ }\textbf {\bibinfo {volume} {126}},\
  \bibinfo {pages} {100402} (\bibinfo {year} {2021})}\BibitemShut {NoStop}%
\bibitem [{\citenamefont {Xue}\ \emph {et~al.}(2022)\citenamefont {Xue},
  \citenamefont {Liu}, \citenamefont {Wang}, \citenamefont {Zhu}, \citenamefont
  {Guo},\ and\ \citenamefont {Wu}}]{Xue2022}%
  \BibitemOpen
  \bibfield  {author} {\bibinfo {author} {\bibfnamefont {S.}~\bibnamefont
  {Xue}}, \bibinfo {author} {\bibfnamefont {Y.}~\bibnamefont {Liu}}, \bibinfo
  {author} {\bibfnamefont {Y.}~\bibnamefont {Wang}}, \bibinfo {author}
  {\bibfnamefont {P.}~\bibnamefont {Zhu}}, \bibinfo {author} {\bibfnamefont
  {C.}~\bibnamefont {Guo}},\ and\ \bibinfo {author} {\bibfnamefont
  {J.}~\bibnamefont {Wu}},\ }\bibfield  {title} {\bibinfo {title} {Variational
  quantum process tomography of unitaries},\ }\href
  {https://doi.org/10.1103/PhysRevA.105.032427} {\bibfield  {journal} {\bibinfo
   {journal} {Phys. Rev. A}\ }\textbf {\bibinfo {volume} {105}},\ \bibinfo
  {pages} {032427} (\bibinfo {year} {2022})}\BibitemShut {NoStop}%
\bibitem [{\citenamefont {Wang}\ \emph
  {et~al.}(2021{\natexlab{a}})\citenamefont {Wang}, \citenamefont {Song},\ and\
  \citenamefont {Wang}}]{Wang2021}%
  \BibitemOpen
  \bibfield  {author} {\bibinfo {author} {\bibfnamefont {X.}~\bibnamefont
  {Wang}}, \bibinfo {author} {\bibfnamefont {Z.}~\bibnamefont {Song}},\ and\
  \bibinfo {author} {\bibfnamefont {Y.}~\bibnamefont {Wang}},\ }\bibfield
  {title} {\bibinfo {title} {Variational quantum singular value
  decomposition},\ }\href {https://doi.org/10.22331/q-2021-06-29-483}
  {\bibfield  {journal} {\bibinfo  {journal} {Quantum}\ }\textbf {\bibinfo
  {volume} {5}},\ \bibinfo {pages} {483} (\bibinfo {year}
  {2021}{\natexlab{a}})}\BibitemShut {NoStop}%
\bibitem [{\citenamefont {Wang}\ \emph {et~al.}(2022)\citenamefont {Wang},
  \citenamefont {Song}, \citenamefont {Zhao}, \citenamefont {Wang},\ and\
  \citenamefont {Wang}}]{Wang2022}%
  \BibitemOpen
  \bibfield  {author} {\bibinfo {author} {\bibfnamefont {K.}~\bibnamefont
  {Wang}}, \bibinfo {author} {\bibfnamefont {Z.}~\bibnamefont {Song}}, \bibinfo
  {author} {\bibfnamefont {X.}~\bibnamefont {Zhao}}, \bibinfo {author}
  {\bibfnamefont {Z.}~\bibnamefont {Wang}},\ and\ \bibinfo {author}
  {\bibfnamefont {X.}~\bibnamefont {Wang}},\ }\bibfield  {title} {\bibinfo
  {title} {Detecting and quantifying entanglement on near-term quantum
  devices},\ }\href {https://doi.org/10.1038/s41534-022-00556-w} {\bibfield
  {journal} {\bibinfo  {journal} {npj Quantum Information}\ }\textbf {\bibinfo
  {volume} {8}},\ \bibinfo {pages} {52} (\bibinfo {year} {2022})}\BibitemShut
  {NoStop}%
\bibitem [{\citenamefont {Mu\~noz Moller}\ \emph {et~al.}(2022)\citenamefont
  {Mu\~noz Moller}, \citenamefont {Pereira}, \citenamefont {Zambrano},
  \citenamefont {Cort\'es-Vega},\ and\ \citenamefont
  {Delgado}}]{munoz2022variational}%
  \BibitemOpen
  \bibfield  {author} {\bibinfo {author} {\bibfnamefont {A.}~\bibnamefont
  {Mu\~noz Moller}}, \bibinfo {author} {\bibfnamefont {L.}~\bibnamefont
  {Pereira}}, \bibinfo {author} {\bibfnamefont {L.}~\bibnamefont {Zambrano}},
  \bibinfo {author} {\bibfnamefont {J.}~\bibnamefont {Cort\'es-Vega}},\ and\
  \bibinfo {author} {\bibfnamefont {A.}~\bibnamefont {Delgado}},\ }\bibfield
  {title} {\bibinfo {title} {Variational determination of multiqubit
  geometrical entanglement in noisy intermediate-scale quantum computers},\
  }\href {https://doi.org/10.1103/PhysRevApplied.18.024048} {\bibfield
  {journal} {\bibinfo  {journal} {Phys. Rev. Applied}\ }\textbf {\bibinfo
  {volume} {18}},\ \bibinfo {pages} {024048} (\bibinfo {year}
  {2022})}\BibitemShut {NoStop}%
\bibitem [{\citenamefont {SHIMONY}(1995)}]{SHIMONY1995}%
  \BibitemOpen
  \bibfield  {author} {\bibinfo {author} {\bibfnamefont {A.}~\bibnamefont
  {SHIMONY}},\ }\bibfield  {title} {\bibinfo {title} {Degree of
  entanglementa},\ }\href {https://doi.org/10.1111/j.1749-6632.1995.tb39008.x}
  {\bibfield  {journal} {\bibinfo  {journal} {Annals of the New York Academy of
  Sciences}\ }\textbf {\bibinfo {volume} {755}},\ \bibinfo {pages} {675}
  (\bibinfo {year} {1995})}\BibitemShut {NoStop}%
\bibitem [{\citenamefont {Barnum}\ and\ \citenamefont
  {Linden}(2001)}]{Barnum2001}%
  \BibitemOpen
  \bibfield  {author} {\bibinfo {author} {\bibfnamefont {H.}~\bibnamefont
  {Barnum}}\ and\ \bibinfo {author} {\bibfnamefont {N.}~\bibnamefont
  {Linden}},\ }\bibfield  {title} {\bibinfo {title} {Monotones and invariants
  for multi-particle quantum states},\ }\href
  {https://doi.org/10.1088/0305-4470/34/35/305} {\bibfield  {journal} {\bibinfo
   {journal} {Journal of Physics A: Mathematical and General}\ }\textbf
  {\bibinfo {volume} {34}},\ \bibinfo {pages} {6787} (\bibinfo {year}
  {2001})}\BibitemShut {NoStop}%
\bibitem [{\citenamefont {Hayashi}\ \emph {et~al.}(2006)\citenamefont
  {Hayashi}, \citenamefont {Markham}, \citenamefont {Murao}, \citenamefont
  {Owari},\ and\ \citenamefont {Virmani}}]{Hayashi2006}%
  \BibitemOpen
  \bibfield  {author} {\bibinfo {author} {\bibfnamefont {M.}~\bibnamefont
  {Hayashi}}, \bibinfo {author} {\bibfnamefont {D.}~\bibnamefont {Markham}},
  \bibinfo {author} {\bibfnamefont {M.}~\bibnamefont {Murao}}, \bibinfo
  {author} {\bibfnamefont {M.}~\bibnamefont {Owari}},\ and\ \bibinfo {author}
  {\bibfnamefont {S.}~\bibnamefont {Virmani}},\ }\bibfield  {title} {\bibinfo
  {title} {Bounds on multipartite entangled orthogonal state discrimination
  using local operations and classical communication},\ }\href
  {https://doi.org/10.1103/PhysRevLett.96.040501} {\bibfield  {journal}
  {\bibinfo  {journal} {Phys. Rev. Lett.}\ }\textbf {\bibinfo {volume} {96}},\
  \bibinfo {pages} {040501} (\bibinfo {year} {2006})}\BibitemShut {NoStop}%
\bibitem [{\citenamefont {Or\'us}\ \emph {et~al.}(2008)\citenamefont {Or\'us},
  \citenamefont {Dusuel},\ and\ \citenamefont {Vidal}}]{Orus2008}%
  \BibitemOpen
  \bibfield  {author} {\bibinfo {author} {\bibfnamefont {R.}~\bibnamefont
  {Or\'us}}, \bibinfo {author} {\bibfnamefont {S.}~\bibnamefont {Dusuel}},\
  and\ \bibinfo {author} {\bibfnamefont {J.}~\bibnamefont {Vidal}},\ }\bibfield
   {title} {\bibinfo {title} {Equivalence of critical scaling laws for
  many-body entanglement in the lipkin-meshkov-glick model},\ }\href
  {https://doi.org/10.1103/PhysRevLett.101.025701} {\bibfield  {journal}
  {\bibinfo  {journal} {Phys. Rev. Lett.}\ }\textbf {\bibinfo {volume} {101}},\
  \bibinfo {pages} {025701} (\bibinfo {year} {2008})}\BibitemShut {NoStop}%
\bibitem [{\citenamefont {Biham}\ \emph {et~al.}(2002)\citenamefont {Biham},
  \citenamefont {Nielsen},\ and\ \citenamefont {Osborne}}]{Biham2002}%
  \BibitemOpen
  \bibfield  {author} {\bibinfo {author} {\bibfnamefont {O.}~\bibnamefont
  {Biham}}, \bibinfo {author} {\bibfnamefont {M.~A.}\ \bibnamefont {Nielsen}},\
  and\ \bibinfo {author} {\bibfnamefont {T.~J.}\ \bibnamefont {Osborne}},\
  }\bibfield  {title} {\bibinfo {title} {Entanglement monotone derived from
  grover's algorithm},\ }\href {https://doi.org/10.1103/PhysRevA.65.062312}
  {\bibfield  {journal} {\bibinfo  {journal} {Phys. Rev. A}\ }\textbf {\bibinfo
  {volume} {65}},\ \bibinfo {pages} {062312} (\bibinfo {year}
  {2002})}\BibitemShut {NoStop}%
\bibitem [{\citenamefont {Wei}\ and\ \citenamefont {Goldbart}(2003)}]{Wei2003}%
  \BibitemOpen
  \bibfield  {author} {\bibinfo {author} {\bibfnamefont {T.-C.}\ \bibnamefont
  {Wei}}\ and\ \bibinfo {author} {\bibfnamefont {P.~M.}\ \bibnamefont
  {Goldbart}},\ }\bibfield  {title} {\bibinfo {title} {Geometric measure of
  entanglement and applications to bipartite and multipartite quantum states},\
  }\href {https://doi.org/10.1103/PhysRevA.68.042307} {\bibfield  {journal}
  {\bibinfo  {journal} {Phys. Rev. A}\ }\textbf {\bibinfo {volume} {68}},\
  \bibinfo {pages} {042307} (\bibinfo {year} {2003})}\BibitemShut {NoStop}%
\bibitem [{\citenamefont {McClean}\ \emph {et~al.}(2018)\citenamefont
  {McClean}, \citenamefont {Boixo}, \citenamefont {Smelyanskiy}, \citenamefont
  {Babbush},\ and\ \citenamefont {Neven}}]{McClean2018}%
  \BibitemOpen
  \bibfield  {author} {\bibinfo {author} {\bibfnamefont {J.~R.}\ \bibnamefont
  {McClean}}, \bibinfo {author} {\bibfnamefont {S.}~\bibnamefont {Boixo}},
  \bibinfo {author} {\bibfnamefont {V.~N.}\ \bibnamefont {Smelyanskiy}},
  \bibinfo {author} {\bibfnamefont {R.}~\bibnamefont {Babbush}},\ and\ \bibinfo
  {author} {\bibfnamefont {H.}~\bibnamefont {Neven}},\ }\bibfield  {title}
  {\bibinfo {title} {Barren plateaus in quantum neural network training
  landscapes},\ }\href {https://doi.org/10.1038/s41467-018-07090-4} {\bibfield
  {journal} {\bibinfo  {journal} {Nature Communications}\ }\textbf {\bibinfo
  {volume} {9}},\ \bibinfo {pages} {4812} (\bibinfo {year} {2018})}\BibitemShut
  {NoStop}%
\bibitem [{\citenamefont {Cerezo}\ \emph
  {et~al.}(2021{\natexlab{b}})\citenamefont {Cerezo}, \citenamefont {Sone},
  \citenamefont {Volkoff}, \citenamefont {Cincio},\ and\ \citenamefont
  {Coles}}]{Cerezo2021BP}%
  \BibitemOpen
  \bibfield  {author} {\bibinfo {author} {\bibfnamefont {M.}~\bibnamefont
  {Cerezo}}, \bibinfo {author} {\bibfnamefont {A.}~\bibnamefont {Sone}},
  \bibinfo {author} {\bibfnamefont {T.}~\bibnamefont {Volkoff}}, \bibinfo
  {author} {\bibfnamefont {L.}~\bibnamefont {Cincio}},\ and\ \bibinfo {author}
  {\bibfnamefont {P.~J.}\ \bibnamefont {Coles}},\ }\bibfield  {title} {\bibinfo
  {title} {Cost function dependent barren plateaus in shallow parametrized
  quantum circuits},\ }\href {https://doi.org/10.1038/s41467-021-21728-w}
  {\bibfield  {journal} {\bibinfo  {journal} {Nature Communications}\ }\textbf
  {\bibinfo {volume} {12}},\ \bibinfo {pages} {1791} (\bibinfo {year}
  {2021}{\natexlab{b}})}\BibitemShut {NoStop}%
\bibitem [{\citenamefont {Wang}\ \emph
  {et~al.}(2021{\natexlab{b}})\citenamefont {Wang}, \citenamefont {Fontana},
  \citenamefont {Cerezo}, \citenamefont {Sharma}, \citenamefont {Sone},
  \citenamefont {Cincio},\ and\ \citenamefont {Coles}}]{Wang2021noise}%
  \BibitemOpen
  \bibfield  {author} {\bibinfo {author} {\bibfnamefont {S.}~\bibnamefont
  {Wang}}, \bibinfo {author} {\bibfnamefont {E.}~\bibnamefont {Fontana}},
  \bibinfo {author} {\bibfnamefont {M.}~\bibnamefont {Cerezo}}, \bibinfo
  {author} {\bibfnamefont {K.}~\bibnamefont {Sharma}}, \bibinfo {author}
  {\bibfnamefont {A.}~\bibnamefont {Sone}}, \bibinfo {author} {\bibfnamefont
  {L.}~\bibnamefont {Cincio}},\ and\ \bibinfo {author} {\bibfnamefont {P.~J.}\
  \bibnamefont {Coles}},\ }\bibfield  {title} {\bibinfo {title} {Noise-induced
  barren plateaus in variational quantum algorithms},\ }\href
  {https://doi.org/10.1038/s41467-021-27045-6} {\bibfield  {journal} {\bibinfo
  {journal} {Nature Communications}\ }\textbf {\bibinfo {volume} {12}},\
  \bibinfo {pages} {6961} (\bibinfo {year} {2021}{\natexlab{b}})}\BibitemShut
  {NoStop}%
\bibitem [{\citenamefont {Khatri}\ \emph {et~al.}(2019)\citenamefont {Khatri},
  \citenamefont {LaRose}, \citenamefont {Poremba}, \citenamefont {Cincio},
  \citenamefont {Sornborger},\ and\ \citenamefont {Coles}}]{Khatri2019}%
  \BibitemOpen
  \bibfield  {author} {\bibinfo {author} {\bibfnamefont {S.}~\bibnamefont
  {Khatri}}, \bibinfo {author} {\bibfnamefont {R.}~\bibnamefont {LaRose}},
  \bibinfo {author} {\bibfnamefont {A.}~\bibnamefont {Poremba}}, \bibinfo
  {author} {\bibfnamefont {L.}~\bibnamefont {Cincio}}, \bibinfo {author}
  {\bibfnamefont {A.~T.}\ \bibnamefont {Sornborger}},\ and\ \bibinfo {author}
  {\bibfnamefont {P.~J.}\ \bibnamefont {Coles}},\ }\bibfield  {title} {\bibinfo
  {title} {Quantum-assisted quantum compiling},\ }\href
  {https://doi.org/10.22331/q-2019-05-13-140} {\bibfield  {journal} {\bibinfo
  {journal} {Quantum}\ }\textbf {\bibinfo {volume} {3}},\ \bibinfo {pages}
  {140} (\bibinfo {year} {2019})}\BibitemShut {NoStop}%
\bibitem [{\citenamefont {Bravo-Prieto}\ \emph {et~al.}(2020)\citenamefont
  {Bravo-Prieto}, \citenamefont {LaRose}, \citenamefont {Cerezo}, \citenamefont
  {Subasi}, \citenamefont {Cincio},\ and\ \citenamefont
  {Coles}}]{BravoPrieto2019}%
  \BibitemOpen
  \bibfield  {author} {\bibinfo {author} {\bibfnamefont {C.}~\bibnamefont
  {Bravo-Prieto}}, \bibinfo {author} {\bibfnamefont {R.}~\bibnamefont
  {LaRose}}, \bibinfo {author} {\bibfnamefont {M.}~\bibnamefont {Cerezo}},
  \bibinfo {author} {\bibfnamefont {Y.}~\bibnamefont {Subasi}}, \bibinfo
  {author} {\bibfnamefont {L.}~\bibnamefont {Cincio}},\ and\ \bibinfo {author}
  {\bibfnamefont {P.~J.}\ \bibnamefont {Coles}},\ }\href
  {https://doi.org/10.48550/ARXIV.1909.05820} {\bibinfo {title} {Variational
  quantum linear solver}} (\bibinfo {year} {2020}),\ \Eprint
  {https://arxiv.org/abs/1909.05820} {arXiv:1909.05820} \BibitemShut {NoStop}%
\bibitem [{\citenamefont {LaRose}\ \emph {et~al.}(2019)\citenamefont {LaRose},
  \citenamefont {Tikku}, \citenamefont {O'Neel-Judy}, \citenamefont {Cincio},\
  and\ \citenamefont {Coles}}]{LaRose2019}%
  \BibitemOpen
  \bibfield  {author} {\bibinfo {author} {\bibfnamefont {R.}~\bibnamefont
  {LaRose}}, \bibinfo {author} {\bibfnamefont {A.}~\bibnamefont {Tikku}},
  \bibinfo {author} {\bibfnamefont {{\'{E}}.}~\bibnamefont {O'Neel-Judy}},
  \bibinfo {author} {\bibfnamefont {L.}~\bibnamefont {Cincio}},\ and\ \bibinfo
  {author} {\bibfnamefont {P.~J.}\ \bibnamefont {Coles}},\ }\bibfield  {title}
  {\bibinfo {title} {Variational quantum state diagonalization},\ }\href
  {https://doi.org/10.1038/s41534-019-0167-6} {\bibfield  {journal} {\bibinfo
  {journal} {npj Quantum Information}\ }\textbf {\bibinfo {volume} {5}},\
  \bibinfo {pages} {57} (\bibinfo {year} {2019})}\BibitemShut {NoStop}%
\bibitem [{\citenamefont {Grant}\ \emph {et~al.}(2019)\citenamefont {Grant},
  \citenamefont {Wossnig}, \citenamefont {Ostaszewski},\ and\ \citenamefont
  {Benedetti}}]{Grant2019initialization}%
  \BibitemOpen
  \bibfield  {author} {\bibinfo {author} {\bibfnamefont {E.}~\bibnamefont
  {Grant}}, \bibinfo {author} {\bibfnamefont {L.}~\bibnamefont {Wossnig}},
  \bibinfo {author} {\bibfnamefont {M.}~\bibnamefont {Ostaszewski}},\ and\
  \bibinfo {author} {\bibfnamefont {M.}~\bibnamefont {Benedetti}},\ }\bibfield
  {title} {\bibinfo {title} {An initialization strategy for addressing barren
  plateaus in parametrized quantum circuits},\ }\href
  {https://doi.org/10.22331/q-2019-12-09-214} {\bibfield  {journal} {\bibinfo
  {journal} {{Quantum}}\ }\textbf {\bibinfo {volume} {3}},\ \bibinfo {pages}
  {214} (\bibinfo {year} {2019})}\BibitemShut {NoStop}%
\bibitem [{\citenamefont {Dborin}\ \emph {et~al.}(2022)\citenamefont {Dborin},
  \citenamefont {Barratt}, \citenamefont {Wimalaweera}, \citenamefont
  {Wright},\ and\ \citenamefont {Green}}]{dborin2022matrix}%
  \BibitemOpen
  \bibfield  {author} {\bibinfo {author} {\bibfnamefont {J.}~\bibnamefont
  {Dborin}}, \bibinfo {author} {\bibfnamefont {F.}~\bibnamefont {Barratt}},
  \bibinfo {author} {\bibfnamefont {V.}~\bibnamefont {Wimalaweera}}, \bibinfo
  {author} {\bibfnamefont {L.}~\bibnamefont {Wright}},\ and\ \bibinfo {author}
  {\bibfnamefont {A.~G.}\ \bibnamefont {Green}},\ }\bibfield  {title} {\bibinfo
  {title} {Matrix product state pre-training for quantum machine learning},\
  }\href {https://doi.org/10.1088/2058-9565/ac7073} {\bibfield  {journal}
  {\bibinfo  {journal} {Quantum Science and Technology}\ }\textbf {\bibinfo
  {volume} {7}},\ \bibinfo {pages} {035014} (\bibinfo {year}
  {2022})}\BibitemShut {NoStop}%
\bibitem [{\citenamefont {Sack}\ \emph {et~al.}(2022)\citenamefont {Sack},
  \citenamefont {Medina}, \citenamefont {Michailidis}, \citenamefont {Kueng},\
  and\ \citenamefont {Serbyn}}]{sack2022avoiding}%
  \BibitemOpen
  \bibfield  {author} {\bibinfo {author} {\bibfnamefont {S.~H.}\ \bibnamefont
  {Sack}}, \bibinfo {author} {\bibfnamefont {R.~A.}\ \bibnamefont {Medina}},
  \bibinfo {author} {\bibfnamefont {A.~A.}\ \bibnamefont {Michailidis}},
  \bibinfo {author} {\bibfnamefont {R.}~\bibnamefont {Kueng}},\ and\ \bibinfo
  {author} {\bibfnamefont {M.}~\bibnamefont {Serbyn}},\ }\bibfield  {title}
  {\bibinfo {title} {Avoiding barren plateaus using classical shadows},\ }\href
  {https://doi.org/10.1103/PRXQuantum.3.020365} {\bibfield  {journal} {\bibinfo
   {journal} {PRX Quantum}\ }\textbf {\bibinfo {volume} {3}},\ \bibinfo {pages}
  {020365} (\bibinfo {year} {2022})}\BibitemShut {NoStop}%
\bibitem [{\citenamefont {Holmes}\ \emph {et~al.}(2022)\citenamefont {Holmes},
  \citenamefont {Sharma}, \citenamefont {Cerezo},\ and\ \citenamefont
  {Coles}}]{holmes2022connecting}%
  \BibitemOpen
  \bibfield  {author} {\bibinfo {author} {\bibfnamefont {Z.}~\bibnamefont
  {Holmes}}, \bibinfo {author} {\bibfnamefont {K.}~\bibnamefont {Sharma}},
  \bibinfo {author} {\bibfnamefont {M.}~\bibnamefont {Cerezo}},\ and\ \bibinfo
  {author} {\bibfnamefont {P.~J.}\ \bibnamefont {Coles}},\ }\bibfield  {title}
  {\bibinfo {title} {Connecting ansatz expressibility to gradient magnitudes
  and barren plateaus},\ }\href {https://doi.org/10.1103/PRXQuantum.3.010313}
  {\bibfield  {journal} {\bibinfo  {journal} {PRX Quantum}\ }\textbf {\bibinfo
  {volume} {3}},\ \bibinfo {pages} {010313} (\bibinfo {year}
  {2022})}\BibitemShut {NoStop}%
\bibitem [{\citenamefont {Wiersema}\ \emph {et~al.}(2021)\citenamefont
  {Wiersema}, \citenamefont {Zhou}, \citenamefont {Carrasquilla},\ and\
  \citenamefont {Kim}}]{wiersema2021measurement}%
  \BibitemOpen
  \bibfield  {author} {\bibinfo {author} {\bibfnamefont {R.}~\bibnamefont
  {Wiersema}}, \bibinfo {author} {\bibfnamefont {C.}~\bibnamefont {Zhou}},
  \bibinfo {author} {\bibfnamefont {J.~F.}\ \bibnamefont {Carrasquilla}},\ and\
  \bibinfo {author} {\bibfnamefont {Y.~B.}\ \bibnamefont {Kim}},\ }\href
  {https://doi.org/10.48550/ARXIV.2111.08035} {\bibinfo {title}
  {Measurement-induced entanglement phase transitions in variational quantum
  circuits}} (\bibinfo {year} {2021}),\ \Eprint
  {https://arxiv.org/abs/2111.08035} {arXiv:2111.08035} \BibitemShut {NoStop}%
\bibitem [{\citenamefont {Kulshrestha}\ and\ \citenamefont
  {Safro}(2022)}]{mele2022avoiding}%
  \BibitemOpen
  \bibfield  {author} {\bibinfo {author} {\bibfnamefont {A.}~\bibnamefont
  {Kulshrestha}}\ and\ \bibinfo {author} {\bibfnamefont {I.}~\bibnamefont
  {Safro}},\ }\href {https://doi.org/10.48550/ARXIV.2204.13751} {\bibinfo
  {title} {Beinit: Avoiding barren plateaus in variational quantum algorithms}}
  (\bibinfo {year} {2022}),\ \Eprint {https://arxiv.org/abs/2204.13751}
  {arXiv:2204.13751} \BibitemShut {NoStop}%
\bibitem [{\citenamefont {Ragone}\ \emph {et~al.}(2022)\citenamefont {Ragone},
  \citenamefont {Braccia}, \citenamefont {Nguyen}, \citenamefont {Schatzki},
  \citenamefont {Coles}, \citenamefont {Sauvage}, \citenamefont {Larocca},\
  and\ \citenamefont {Cerezo}}]{Ragone2022}%
  \BibitemOpen
  \bibfield  {author} {\bibinfo {author} {\bibfnamefont {M.}~\bibnamefont
  {Ragone}}, \bibinfo {author} {\bibfnamefont {P.}~\bibnamefont {Braccia}},
  \bibinfo {author} {\bibfnamefont {Q.~T.}\ \bibnamefont {Nguyen}}, \bibinfo
  {author} {\bibfnamefont {L.}~\bibnamefont {Schatzki}}, \bibinfo {author}
  {\bibfnamefont {P.~J.}\ \bibnamefont {Coles}}, \bibinfo {author}
  {\bibfnamefont {F.}~\bibnamefont {Sauvage}}, \bibinfo {author} {\bibfnamefont
  {M.}~\bibnamefont {Larocca}},\ and\ \bibinfo {author} {\bibfnamefont
  {M.}~\bibnamefont {Cerezo}},\ }\href@noop {} {\bibinfo {title}
  {Representation theory for geometric quantum machine learning}} (\bibinfo
  {year} {2022}),\ \Eprint {https://arxiv.org/abs/arXiv:2210.07980}
  {arXiv:2210.07980} \BibitemShut {NoStop}%
\bibitem [{IBM(2021)}]{IBMQ}%
  \BibitemOpen
  \href@noop {} {\bibinfo {title} {Ibm quantum}},\ \bibinfo {howpublished}
  {\url{https://quantum-computing.ibm.com/}} (\bibinfo {year}
  {2021})\BibitemShut {NoStop}%
\bibitem [{\citenamefont {Spall}(1992)}]{spallspsa1992}%
  \BibitemOpen
  \bibfield  {author} {\bibinfo {author} {\bibfnamefont {J.}~\bibnamefont
  {Spall}},\ }\bibfield  {title} {\bibinfo {title} {Multivariate stochastic
  approximation using a simultaneous perturbation gradient approximation},\
  }\href {https://doi.org/10.1109/9.119632} {\bibfield  {journal} {\bibinfo
  {journal} {IEEE Transactions on Automatic Control}\ }\textbf {\bibinfo
  {volume} {37}},\ \bibinfo {pages} {332} (\bibinfo {year} {1992})}\BibitemShut
  {NoStop}%
\bibitem [{\citenamefont {Gross}\ \emph {et~al.}(2009)\citenamefont {Gross},
  \citenamefont {Flammia},\ and\ \citenamefont {Eisert}}]{Most_Gross}%
  \BibitemOpen
  \bibfield  {author} {\bibinfo {author} {\bibfnamefont {D.}~\bibnamefont
  {Gross}}, \bibinfo {author} {\bibfnamefont {S.~T.}\ \bibnamefont {Flammia}},\
  and\ \bibinfo {author} {\bibfnamefont {J.}~\bibnamefont {Eisert}},\
  }\bibfield  {title} {\bibinfo {title} {Most quantum states are too entangled
  to be useful as computational resources},\ }\href
  {https://doi.org/10.1103/PhysRevLett.102.190501} {\bibfield  {journal}
  {\bibinfo  {journal} {Phys. Rev. Lett.}\ }\textbf {\bibinfo {volume} {102}},\
  \bibinfo {pages} {190501} (\bibinfo {year} {2009})}\BibitemShut {NoStop}%
\bibitem [{\citenamefont {Wales}\ and\ \citenamefont {Doye}(1997)}]{bashop}%
  \BibitemOpen
  \bibfield  {author} {\bibinfo {author} {\bibfnamefont {D.~J.}\ \bibnamefont
  {Wales}}\ and\ \bibinfo {author} {\bibfnamefont {J.~P.~K.}\ \bibnamefont
  {Doye}},\ }\bibfield  {title} {\bibinfo {title} {Global optimization by
  {B}asin-{H}opping and the lowest energy structures of {L}ennard-{J}ones
  clusters containing up to 110 atoms},\ }\href
  {https://doi.org/10.1021/jp970984n} {\bibfield  {journal} {\bibinfo
  {journal} {J. Phys. Chem. A}\ }\textbf {\bibinfo {volume} {101}},\ \bibinfo
  {pages} {5111} (\bibinfo {year} {1997})}\BibitemShut {NoStop}%
\bibitem [{\citenamefont {Hayashi}\ \emph {et~al.}(2009)\citenamefont
  {Hayashi}, \citenamefont {Markham}, \citenamefont {Murao}, \citenamefont
  {Owari},\ and\ \citenamefont {Virmani}}]{suma_ghz_w}%
  \BibitemOpen
  \bibfield  {author} {\bibinfo {author} {\bibfnamefont {M.}~\bibnamefont
  {Hayashi}}, \bibinfo {author} {\bibfnamefont {D.}~\bibnamefont {Markham}},
  \bibinfo {author} {\bibfnamefont {M.}~\bibnamefont {Murao}}, \bibinfo
  {author} {\bibfnamefont {M.}~\bibnamefont {Owari}},\ and\ \bibinfo {author}
  {\bibfnamefont {S.}~\bibnamefont {Virmani}},\ }\bibfield  {title} {\bibinfo
  {title} {The geometric measure of entanglement for a symmetric pure state
  with non-negative amplitudes},\ }\href {https://doi.org/10.1063/1.3271041}
  {\bibfield  {journal} {\bibinfo  {journal} {Journal of Mathematical Physics}\
  }\textbf {\bibinfo {volume} {50}},\ \bibinfo {pages} {122104} (\bibinfo
  {year} {2009})}\BibitemShut {NoStop}%
\bibitem [{\citenamefont {Hradil}(1997)}]{Hradil1997_MLE}%
  \BibitemOpen
  \bibfield  {author} {\bibinfo {author} {\bibfnamefont {Z.}~\bibnamefont
  {Hradil}},\ }\bibfield  {title} {\bibinfo {title} {Quantum-state
  estimation},\ }\href {https://doi.org/10.1103/PhysRevA.55.R1561} {\bibfield
  {journal} {\bibinfo  {journal} {Phys. Rev. A}\ }\textbf {\bibinfo {volume}
  {55}},\ \bibinfo {pages} {R1561} (\bibinfo {year} {1997})}\BibitemShut
  {NoStop}%
\bibitem [{\citenamefont {Shang}\ \emph {et~al.}(2017)\citenamefont {Shang},
  \citenamefont {Zhang},\ and\ \citenamefont {Ng}}]{Shang2017_superfast}%
  \BibitemOpen
  \bibfield  {author} {\bibinfo {author} {\bibfnamefont {J.}~\bibnamefont
  {Shang}}, \bibinfo {author} {\bibfnamefont {Z.}~\bibnamefont {Zhang}},\ and\
  \bibinfo {author} {\bibfnamefont {H.~K.}\ \bibnamefont {Ng}},\ }\bibfield
  {title} {\bibinfo {title} {Superfast maximum-likelihood reconstruction for
  quantum tomography},\ }\href {https://doi.org/10.1103/PhysRevA.95.062336}
  {\bibfield  {journal} {\bibinfo  {journal} {Phys. Rev. A}\ }\textbf {\bibinfo
  {volume} {95}},\ \bibinfo {pages} {062336} (\bibinfo {year}
  {2017})}\BibitemShut {NoStop}%
\bibitem [{\citenamefont {Bravyi}\ \emph {et~al.}(2021)\citenamefont {Bravyi},
  \citenamefont {Sheldon}, \citenamefont {Kandala}, \citenamefont {Mckay},\
  and\ \citenamefont {Gambetta}}]{Bravyi2021_errormitigation}%
  \BibitemOpen
  \bibfield  {author} {\bibinfo {author} {\bibfnamefont {S.}~\bibnamefont
  {Bravyi}}, \bibinfo {author} {\bibfnamefont {S.}~\bibnamefont {Sheldon}},
  \bibinfo {author} {\bibfnamefont {A.}~\bibnamefont {Kandala}}, \bibinfo
  {author} {\bibfnamefont {D.~C.}\ \bibnamefont {Mckay}},\ and\ \bibinfo
  {author} {\bibfnamefont {J.~M.}\ \bibnamefont {Gambetta}},\ }\bibfield
  {title} {\bibinfo {title} {Mitigating measurement errors in multiqubit
  experiments},\ }\href {https://doi.org/10.1103/PhysRevA.103.042605}
  {\bibfield  {journal} {\bibinfo  {journal} {Phys. Rev. A}\ }\textbf {\bibinfo
  {volume} {103}},\ \bibinfo {pages} {042605} (\bibinfo {year}
  {2021})}\BibitemShut {NoStop}%
\bibitem [{\citenamefont {Mooney}\ \emph
  {et~al.}(2021{\natexlab{a}})\citenamefont {Mooney}, \citenamefont {White},
  \citenamefont {Hill},\ and\ \citenamefont {Hollenberg}}]{Mooney_2021}%
  \BibitemOpen
  \bibfield  {author} {\bibinfo {author} {\bibfnamefont {G.~J.}\ \bibnamefont
  {Mooney}}, \bibinfo {author} {\bibfnamefont {G.~A.~L.}\ \bibnamefont
  {White}}, \bibinfo {author} {\bibfnamefont {C.~D.}\ \bibnamefont {Hill}},\
  and\ \bibinfo {author} {\bibfnamefont {L.~C.~L.}\ \bibnamefont
  {Hollenberg}},\ }\bibfield  {title} {\bibinfo {title} {Generation and
  verification of 27-qubit greenberger-horne-zeilinger states in a
  superconducting quantum computer},\ }\href
  {https://doi.org/10.1088/2399-6528/ac1df7} {\bibfield  {journal} {\bibinfo
  {journal} {Journal of Physics Communications}\ }\textbf {\bibinfo {volume}
  {5}},\ \bibinfo {pages} {095004} (\bibinfo {year}
  {2021}{\natexlab{a}})}\BibitemShut {NoStop}%
\bibitem [{\citenamefont {Mooney}\ \emph
  {et~al.}(2021{\natexlab{b}})\citenamefont {Mooney}, \citenamefont {White},
  \citenamefont {Hill},\ and\ \citenamefont {Hollenberg}}]{Mooney_2021_2}%
  \BibitemOpen
  \bibfield  {author} {\bibinfo {author} {\bibfnamefont {G.~J.}\ \bibnamefont
  {Mooney}}, \bibinfo {author} {\bibfnamefont {G.~A.~L.}\ \bibnamefont
  {White}}, \bibinfo {author} {\bibfnamefont {C.~D.}\ \bibnamefont {Hill}},\
  and\ \bibinfo {author} {\bibfnamefont {L.~C.~L.}\ \bibnamefont
  {Hollenberg}},\ }\bibfield  {title} {\bibinfo {title} {Whole-device
  entanglement in a 65-qubit superconducting quantum computer},\ }\href
  {https://doi.org/https://doi.org/10.1002/qute.202100061} {\bibfield
  {journal} {\bibinfo  {journal} {Advanced Quantum Technologies}\ }\textbf
  {\bibinfo {volume} {4}},\ \bibinfo {pages} {2100061} (\bibinfo {year}
  {2021}{\natexlab{b}})}\BibitemShut {NoStop}%
\bibitem [{\citenamefont {Utreras-Alarc{\'o}n}\ \emph
  {et~al.}(2019)\citenamefont {Utreras-Alarc{\'o}n}, \citenamefont
  {Rivera-Tapia}, \citenamefont {Niklitschek},\ and\ \citenamefont
  {Delgado}}]{URND2019}%
  \BibitemOpen
  \bibfield  {author} {\bibinfo {author} {\bibfnamefont {A.}~\bibnamefont
  {Utreras-Alarc{\'o}n}}, \bibinfo {author} {\bibfnamefont {M.}~\bibnamefont
  {Rivera-Tapia}}, \bibinfo {author} {\bibfnamefont {S.}~\bibnamefont
  {Niklitschek}},\ and\ \bibinfo {author} {\bibfnamefont {A.}~\bibnamefont
  {Delgado}},\ }\bibfield  {title} {\bibinfo {title} {Stochastic optimization
  on complex variables and pure-state quantum tomography},\ }\href
  {https://doi.org/10.1038/s41598-019-52289-0} {\bibfield  {journal} {\bibinfo
  {journal} {Sci. Rep.}\ }\textbf {\bibinfo {volume} {9}},\ \bibinfo {pages}
  {16143} (\bibinfo {year} {2019})}\BibitemShut {NoStop}%
\bibitem [{\citenamefont {Kushner}\ and\ \citenamefont
  {Clark}(1978)}]{Kushner1978}%
  \BibitemOpen
  \bibfield  {author} {\bibinfo {author} {\bibfnamefont {H.~J.}\ \bibnamefont
  {Kushner}}\ and\ \bibinfo {author} {\bibfnamefont {D.~S.}\ \bibnamefont
  {Clark}},\ }\href {https://doi.org/10.1007/978-1-4684-9352-8} {\emph
  {\bibinfo {title} {Stochastic Approximation Methods for Constrained and
  Unconstrained Systems}}},\ Vol.~\bibinfo {volume} {26}\ (\bibinfo
  {publisher} {Springer New York},\ \bibinfo {year} {1978})\BibitemShut
  {NoStop}%
\bibitem [{\citenamefont {Kushner}\ and\ \citenamefont
  {Yin}(1997)}]{Kushner1997}%
  \BibitemOpen
  \bibfield  {author} {\bibinfo {author} {\bibfnamefont {H.~J.}\ \bibnamefont
  {Kushner}}\ and\ \bibinfo {author} {\bibfnamefont {G.~G.}\ \bibnamefont
  {Yin}},\ }\href {https://doi.org/10.1007/978-1-4899-2696-8} {\emph {\bibinfo
  {title} {Stochastic Approximation Algorithms and Applications}}}\ (\bibinfo
  {publisher} {Springer New York},\ \bibinfo {year} {1997})\BibitemShut
  {NoStop}%
\bibitem [{\citenamefont {Kushner}\ and\ \citenamefont
  {Yin}(2003)}]{Kushner2003}%
  \BibitemOpen
  \bibfield  {author} {\bibinfo {author} {\bibfnamefont {H.}~\bibnamefont
  {Kushner}}\ and\ \bibinfo {author} {\bibfnamefont {G.}~\bibnamefont {Yin}},\
  }\href {https://doi.org/10.1007/b97441} {\emph {\bibinfo {title} {Stochastic
  Approximation and Recursive Algorithms and Applications}}},\ Vol.~\bibinfo
  {volume} {35}\ (\bibinfo  {publisher} {Springer-Verlag},\ \bibinfo {year}
  {2003})\BibitemShut {NoStop}%
\bibitem [{\citenamefont {Spall}(2007)}]{Spall2007}%
  \BibitemOpen
  \bibfield  {author} {\bibinfo {author} {\bibfnamefont {J.}~\bibnamefont
  {Spall}},\ }\bibfield  {title} {\bibinfo {title} {Introduction to stochastic
  search and optimization. estimation, simulation, and control},\ }\href
  {https://doi.org/10.1002/0471722138} {\bibfield  {journal} {\bibinfo
  {journal} {IEEE Transactions on Neural Networks}\ }\textbf {\bibinfo {volume}
  {18}} (\bibinfo {year} {2007})}\BibitemShut {NoStop}%
\bibitem [{\citenamefont {Albert}\ and\ \citenamefont
  {Gardner}(2003)}]{Albert2003}%
  \BibitemOpen
  \bibfield  {author} {\bibinfo {author} {\bibfnamefont {A.~E.}\ \bibnamefont
  {Albert}}\ and\ \bibinfo {author} {\bibfnamefont {L.~A.}\ \bibnamefont
  {Gardner}},\ }\href {https://doi.org/10.7551/mitpress/6464.001.0001} {\emph
  {\bibinfo {title} {Stochastic Approximation and {NonLinear} Regression}}}\
  (\bibinfo  {publisher} {The {MIT} Press},\ \bibinfo {year}
  {2003})\BibitemShut {NoStop}%
\bibitem [{\citenamefont {Bhatnagar}\ \emph {et~al.}(2013)\citenamefont
  {Bhatnagar}, \citenamefont {Prasad},\ and\ \citenamefont
  {Prashanth}}]{Bhatnagar2013}%
  \BibitemOpen
  \bibfield  {author} {\bibinfo {author} {\bibfnamefont {S.}~\bibnamefont
  {Bhatnagar}}, \bibinfo {author} {\bibfnamefont {H.}~\bibnamefont {Prasad}},\
  and\ \bibinfo {author} {\bibfnamefont {L.}~\bibnamefont {Prashanth}},\ }\href
  {https://doi.org/10.1007/978-1-4471-4285-0} {\emph {\bibinfo {title}
  {Stochastic Recursive Algorithms for Optimization}}},\ Vol.\ \bibinfo
  {volume} {434}\ (\bibinfo  {publisher} {Springer London},\ \bibinfo {year}
  {2013})\BibitemShut {NoStop}%
\bibitem [{\citenamefont {Wirtinger}(1927)}]{Wirtinger1927}%
  \BibitemOpen
  \bibfield  {author} {\bibinfo {author} {\bibfnamefont {W.}~\bibnamefont
  {Wirtinger}},\ }\bibfield  {title} {\bibinfo {title} {Zur formalen theorie
  der funktionen von mehr komplexen veränderlichen},\ }\href
  {http://eudml.org/doc/182642} {\bibfield  {journal} {\bibinfo  {journal}
  {Mathematische Annalen}\ }\textbf {\bibinfo {volume} {97}},\ \bibinfo {pages}
  {357} (\bibinfo {year} {1927})}\BibitemShut {NoStop}%
\bibitem [{\citenamefont {Kreutz-Delgado}(2009)}]{Kreutz-Delgado2009}%
  \BibitemOpen
  \bibfield  {author} {\bibinfo {author} {\bibfnamefont {K.}~\bibnamefont
  {Kreutz-Delgado}},\ }\href@noop {} {\bibinfo {title} {The complex gradient
  operator and the {CR}-calculus}} (\bibinfo {year} {2009}),\ \Eprint
  {https://arxiv.org/abs/0906.4835} {arXiv:0906.4835 [math.OC]} \BibitemShut
  {NoStop}%
\end{thebibliography}%

\appendix


\section{\label{sec: Method} Bounds for the Infidelity}
The aim of this section is to obtain bounds for the global infidelity Eq.~\eqref{inf=gme}. Let us consider a $n$-qubit Hamiltonian $H$ with eigenstates $\{|E_k\rangle\}$ and possibly degenerate eigenvalues $\{E_k\}$ with $k=0,\dots,2^n-1$. The eigenvalues are ordered in increasing order and $E_0=0$. For an arbitrary density matrix $\rho$, the energy expectation value is given by
\begin{equation}
    \mathrm{Tr}(H\rho) = \sum_{k=0}^{2^n-1} E_k \langle E_k | \rho | E_k \rangle.
\end{equation}
Removing the ground state from the sum,
\begin{equation}
    \mathrm{Tr}(H\rho) = \sum_{k=1}^{2^n-1} E_k \langle E_k | \rho | E_k \rangle,
\end{equation}
and considering that $E_1\leq E_k$ for $k\geq 1$, we have that
\begin{equation}
    \mathrm{Tr}( H\rho ) \geq E_1 \sum_{k=1}^{2^n-1} \langle E_k | \rho | E_k \rangle.
\end{equation}
Due to the normalization of $\rho$, that is,  
\begin{equation}
\sum_{k=0}^{2^n-1} \langle E_k | \rho | E_k \rangle = 1,    
\end{equation}
we obtain 
\begin{equation}
    \mathrm{Tr}( H\rho ) \geq E_1 ( 1 - \langle E_0 | \rho | E_0 \rangle ).
\end{equation}
Isolating the fidelity between $\rho$ and $|E_0 \rangle\langle E_0 |$, we achieve a lower bound for the fidelity between the ground state and an arbitrary state $\rho$:
\begin{equation}\label{eq: LB to the fidelity}
\langle E_0 | \rho | E_0 \rangle \geq  1- \frac{1}{E_1}\mathrm{Tr}(\rho H).     
\end{equation}
Furthermore, by using a similar reasoning involving the highest eigenvalue $E_{max}$, we obtain
\begin{equation}\label{eq: UB to the fidelity}
\begin{split}
\langle E_0 | \rho | E_0 \rangle \leq  1-\frac{1}{E_{max}} \mathrm{Tr}(H \rho ).
\end{split}    
\end{equation}
This means that by finding a suitable Hamiltonian $H$ we can get the upper \eqref{eq: LB to the fidelity} and lower \eqref{eq: UB to the fidelity} bounds for the fidelity between the ground state $|E_0\rangle$ and an arbitrary state $\rho$. In the following, we construct a $n$-qubit Hamiltonian with ground state $|E_0 \rangle = |0_1 \rangle \otimes \cdots \otimes |0_n\rangle$ that provides bounds for the GME.

To construct the Hamiltonian, we define the operators
\begin{align}
    \Pi_{i j} =  |0_i \rangle \langle 0_{i}| \otimes |0_j \rangle \langle 0_j|,
\end{align} 
with $1 \leq i, j \leq n$ and $i \neq j$. These are projectors that act non-trivially only on qubits $i$ and $j$ and satisfy $\Pi_{ij} = \Pi_{ji}$. 
We now define the operator $\Gamma$ as the sum of all these $n(n-1)$ local projectors, that is,
\begin{align}
    \Gamma = \frac{1}{2}\sum_{\substack{i, j =1 \\ i \neq j}}^n \Pi_{ij} = \sum_{\substack{i, j =1 \\ i < j}}^n \Pi_{ij}. 
\end{align}
The diagonal decomposition of the above operator is given by
\begin{align}\label{eq:Gamma}
    \Gamma = \sum_{k = 0}^{n-2} \frac{(n-k)(n-k-1)}{2} P(k),
\end{align}
where $P(k) =\sum_{|\alpha\rangle \in \lambda_k} | \alpha \rangle\langle \alpha |$ is the projector over the set $\lambda_k$ of states in the computational basis containing $k$ qubits each in the state $|1 \rangle$. For example, $P(0)$ projects onto the subspace defined by the state $| \Phi_0 \rangle = \bigotimes_{i=1}^n  |0_i\rangle$. This state is an eigenvector of every local operator $\Pi_{ij}$ with an eigenvalue equal to $1$. Therefore, $| \Phi_0 \rangle$ is also eigenvector of $\Gamma$ with eigenvalue ${n(n-1)}/{2}$. The next projector, $P(1)$, is associated to the states $ \left( \bigotimes_{i \neq k}  |0_i\rangle \right) \otimes |1_k\rangle$, for $k=1,\dots, n$. Each one of these states is an eigenvector of operators $\Pi_{k, i}$ and $\Pi_{i, k}$ with eigenvalue equal to $0$, and eigenvector of the remaining $n(n-1)-(2n-2)$ operators $\Pi_{i \neq k, j \neq k}$ with eigenvalue equal to $1$. Then, they all have eigenvalue ${(n-1)(n-2)}/{2}$ of $\Gamma$.

We now use $\Gamma$ to define the Hamiltonian 
\begin{equation}
    H_L = \mathds{1} - \frac{2}{n(n-1)} \Gamma
\end{equation}
or equivalently
\begin{equation}
    H_L = \mathds{1} - \sum_{k = 0}^{n-2} \frac{(n-k)(n-k-1)}{n(n-1)} P(k).
\end{equation}

The lowest eigenvalue of $H_L$ is $E_0 = 0$, associated to $P(0) = | \Phi_0 \rangle \langle \Phi_0 |$. The next eigenvalue is $E_1 =\frac{2}{n}$, associated to $P(1)$. Meanwhile, the highest eigenvalue is $E_{max} = 1$, associated to $P(n-1)$ and $P(n)$. Therefore, this Hamiltonian satisfies the requirements of the inequalities given by Eq.~\eqref{eq: LB to the fidelity} and Eq.~\eqref{eq: UB to the fidelity}, obtaining the following  bounds for the fidelity in terms of the expectation value of $H_L$ 
\begin{align}\label{eq:fidelity_bounds}
     1-\mathrm{Tr}(H_L \rho ) \geq \langle \Phi_0 | \rho | \Phi_0 \rangle \geq  1- \frac{n}{2}\mathrm{Tr}(\rho H_L). 
\end{align}
This means that by minimizing the expectation value of $H_L$, or equivalently, maximizing the expectation value of $\Gamma$, we obtain a stricter bound for the fidelity. 

Equation \ref{eq:fidelity_bounds} is enough to motivate our proposal. However, these bounds may not be tight or may even be trivial for some states. In the general case, both Hamiltonians $H_L$ and $H_G$ have the same ground state $|0\rangle^{\otimes n}$. Then, following the variational principle, they both should converge to the same ground state.  

\section{Evaluation of the expectation value of $H_L$}
\label{sec: Sampling}

In this section, we show how to evaluate the expectation value of the Hamiltonian $H_L$ \eqref{eq: local hamiltonian} using the Monte Carlo method. First, let us notice that the Hamiltonian can be rewritten as
\begin{equation}
    H_L = \frac{1}{n(n-1)}\sum_{\substack{i, j =1 \\ i \neq j}}^n \left( \mathds{1}  - \Pi_{ij} \right).
\end{equation}
Thereby, the expectation value $ \langle  H_L  \rangle_{\theta} := \langle \alpha (\boldsymbol{\theta})|  H_L | \alpha (\boldsymbol{\theta})\rangle=\mathrm{Tr}(\rho H_L)$ is given by
\begin{equation}\label{hamiltonian_mc}
    \langle  H_L  \rangle_{\theta} = \frac{1}{n(n-1)}\sum_{\substack{i, j =1 \\ i \neq j}}^n I_{ij},
\end{equation}
where $I_{ij}$ are the two-qubit local infidelities
\begin{equation}
    I_{ij} =1 - \mathrm{Tr}( \rho \Pi_{ij}),
\end{equation}
with $\rho =  | \alpha (\boldsymbol{\theta}) \rangle \langle \alpha (\boldsymbol{\theta})|$. The infidelities satisfy $I_{ij} = I_{ji}$. To estimate $\langle  H_L  \rangle_{\theta}$, we use the random variable 
\begin{align*}
X_g & = \frac{1}{\lfloor n/2 \rfloor}\sum_{(i,j)\in g} I_{ij} (\boldsymbol{\theta}_i,\boldsymbol{\theta}_j)\\
& = \frac{1}{\lfloor n/2 \rfloor} \sum_{k=1}^{\lfloor n/2 \rfloor} X^{(k)} (\boldsymbol{\theta}_i,\boldsymbol{\theta}_j), 
\end{align*}
where $g\in \mathcal{G}$ is a uniformly randomly selected non-overlapping set of pairs. Notice that $X^{(k)}$ is also a random variable whose value corresponds to the local infidelity $I_{i_k j_k}$ of the subsystem $(i_k, j_k)$ of qubits, defined by the $k$-th pair of the set $g$. Moreover, the random variables $X^{(k)}$ are correlated since the pairs $(i_k,j_k)$ appearing in each $X^{(k)}$ are sequentially taken to define a non-overlapping set of pairs. Namely, the $k$-th pair $(i_{k},j_{k})$ is randomly selected on the remaining set of $\{1,n\}$ after removing all the pairs $(i_{k'},j_{k'})$ with $k'<k$.


The performance of $X_g$ on the estimation of $\langle H_L \rangle_{\theta}$ is measured by the mean squared error ($\textrm{MSE}$). To calculate this function, we need to compute the marginal probabilities $\mathbb{P}(X^{(k)} = I_{ij})$ and the joint probabilities $\mathbb{P}(X^{(k)} = I_{ij}, X^{(k')} = I_{i'j'})$. First, we note that for a given $X^{(k)}$ and a fixed pair $(i,j)$, the number of non-overlappings set of pairs in $\mathcal{G}$ such that $X^{(k)} = I_{ij}$ is $(n-2)!$, and the total number of pair orderings $\vert \mathcal{G} \vert$ is $n!$.  Then, for any $k=1,\dots, \lfloor n/2\rfloor$,
\begin{align*}
    \mathbb{P}(X^{(k)} = I_{ij}) =   \frac{(n-2)!}{n!} = \frac{1}{n(n-1)}.
\end{align*}
Therefore, all $X^{(k)}$ follow the same probability distribution, with mean value $\mathbb{E}\left(X^{(k)}\right) = \langle H_L \rangle_{\theta}$. Furthermore, by an analogous computation, we obtain  
\begin{align*}
   \mathbb{P}(X^{(k)} = I_{ij}, X^{(k')} = I_{i'j'}) = \frac{1}{n(n-1)(n-2)(n-3)},
\end{align*}
for any $i \neq j \neq i' \neq j'$, and zero otherwise. \\
From the previous calculation, the expectation value of $X_g$ is given by 
\begin{equation}\label{eq: EV X_g}
\begin{split}
\mathbb{E} \left(X_g\right)  & =   \frac{1}{\lfloor n/2 \rfloor} \sum_{k=1}^{\lfloor n/2\rfloor} \mathbb{E} \left( X^{(k)} \right) \\
& = \frac{1}{\lfloor n/2 \rfloor} \sum_{k=1}^{\lfloor n/2\rfloor} \sum_{\substack{i, j =1 \\ i \neq j}}^n \frac{1}{n(n-1)} I_{ij}  \\
& =  \sum_{\substack{i, j =1 \\ i \neq j}}^n \frac{1}{n(n-1)} I_{ij}\\
& =  \langle H_L \rangle_{\theta}.
\end{split}    
\end{equation}
Consequently $X_g$ is an unbiased estimator of $\langle H_L \rangle_{\theta}$, and from this, we conclude that
\begin{equation*}
\textrm{MSE}(\langle  H_L  \rangle_{\theta},X_g) = \mathbb{E}\left((X_g - \langle  H_L  \rangle_{\theta} )^2\right) = \textrm{Var}(X_g).
\end{equation*}
To calculate the variance of $X_g$, we note that
\begin{equation}\label{eq: JointEV}
\hspace{-0.2 cm}
\begin{split}
&\mathbb{E}\left(X^{(k)} X^{(k')}\right) \\
& = \sum_{i\neq j} \sum_{\substack{i'\neq j'\\ i',j' \notin \{ i,j \} }} 
\frac{1}{n (n-1)(n-2)(n-3)} I_{i j} I_{ i' j'} \\
& \leq \frac{2n(n-1)}{(n-2)(n-3)} \langle  H_L  \rangle_{\theta}^2. 
\end{split}
\end{equation}
Therefore, from $\mathbb{E} ((X^{(k)} - \langle  H_L  \rangle_{\theta} )^2 ) \leq 1$ together with equations \eqref{eq: EV X_g} and \eqref{eq: JointEV}, we get
\begin{equation}
\begin{split}
& \textrm{Var}(X_g)  = \mathbb{E}\left(\left(X_g - \langle  H_L  \rangle_{\theta}\right)^2\right)\\
& \quad = \mathbb{E}\left( \left(\frac{2}{n} \sum_{k = 1}^{n/2}  \left(X_k - \langle  H_L  \rangle_{\theta} \right) \right)^2 \right)\\
& \quad = \frac{4}{n^2} \left\{ \sum_{k = 1}^{n/2} \mathbb{E} \left(\left(X_k - \langle  H_L  \rangle_{\theta} \right)^2\right) \right. \\
& \qquad  \left. +  \sum_{k,k' = 1}^{n/2} \mathbb{E} \left( \left(X_k - \langle  H_L  \rangle_{\theta} \right) \left(X_{k'} - \langle  H_L  \rangle_{\theta} \right) \right) \right\}\\
& \quad \leq \frac{2}{n} +  \left(\frac{(n-1)}{(n-2)(n-3)} - \frac{2}{n} \right) \langle  H_L  \rangle_{\theta}^2.
\end{split}    
\end{equation}
This means that the mean square error can be upper-bounded as
\begin{equation}
    \textrm{MSE}\left(\langle  H_L  \rangle_{\theta},X_g \right) \leq \frac{2}{n} +  \left(\frac{(n-1)}{(n-2)(n-3)} - \frac{2}{n} \right) \langle  H_L  \rangle_{\theta}^2.  
\end{equation}
The above procedure can be implemented efficiently by simultaneously measuring all local infidelities corresponding to a set of non-overlapping pairs of qubits.

\section{\label{sec:A}Calculation of exact GME value}

In order to benchmark the results of iVDGE and VDGE we need the exact value of the GME for different states. Unfortunately, for arbitrary states this must be done numerically. In Fig. \ref{fig:Fig1} we parameterize $n$ separable states $\{ |\Phi_i \rangle =  \cos \theta_i |0 \rangle + e^{i \phi_i} \sin \theta_i  |1 \rangle \}$ and minimize the function 
\begin{align}\label{ansatz_bh}
    1 - \left| \langle \Psi | \left( |\Phi_1\rangle \otimes \dots \otimes |\Phi_n\rangle  \right) \right|^2
\end{align}
via Basin-hopping optimization algorithm. In this case, Eq. \eqref{ansatz_bh} is calculated exactly. We repeat this procedure 20 times and keep the lowest value.

For a high number of qubits, the previous procedure is not reliable, so we have to restrict our attention to symmetric states, whose GME is optimized by a symmetric separable state \cite{suma_ghz_w}. Then, the ansatz is  $|\Phi \rangle \otimes \dots \otimes |\Phi \rangle$, with $|\Phi\rangle = \cos \theta |0 \rangle + \sin \theta e^{i \phi} |1 \rangle$. This means that the ansatz only depends on two parameters, greatly reducing the search space. We use this in Fig. \ref{fig:Fig2}, where we minimize 
\begin{align}
    1 - \left| \langle \Psi | \left( |\Phi\rangle \otimes \dots \otimes |\Phi\rangle  \right) \right|^2
\end{align}
via Basin-hopping optimization algorithm, repeating the procedure $20$ times and keeping the lowest obtained GME value.

\section{\label{sec:CSPSA} CSPSA algorithm}

The complex simultaneous perturbation stochastic approximation (CSPSA) method~\cite{URND2019} is an stochastic optimisation algorithm \cite{Kushner1978,Kushner1997,Kushner2003,Spall2007,Albert2003,Bhatnagar2013}, used to find a global minimum/maximum of a given real-valued function $f$ of complex variables. Since non-constant real-valued functions of complex arguments violate the Cauchy-Riemann conditions, optimization methods based on its derivatives are not functional. CSPSA avoid this problem by a stochastic approximation of the gradient $\nabla f$ computed by two evaluations of the target function in a similar way to the calculation of a central finite difference. CAPSA is based on Wirtinger's calculus \cite{Wirtinger1927,Kreutz-Delgado2009} where a function $f$ depends on a complex variable $z$ and its conjugate $z^*$, which are considered independent, i.e. $f(z)=f(z,z^*)$. Stationary points of $f$ are completely characterized by the vanishing of the gradient $\partial_{z^*}f = 0$ or equivalently, by $\partial_{z} f = 0$. Thereby, the CSPSA stochastic gradient approximation on the $k$-th iteration is computed by
\begin{align}\label{eq:CSPSA Gradient}
  \nabla f (\bm\theta) \approx \frac{f(\bm\theta + c_{k}\bm\Delta_{k}) - f(\bm\theta - c_{k}\bm\Delta_{k})}{2c_{k}}
  \begin{pmatrix}
    1/\Delta_{k,1} \\
    \vdots \\
    1/\Delta_{k,d}
  \end{pmatrix},
\end{align}
where $\bm\Delta_{k}$ is a random perturbation vector with $d$ components uniformly generated from the set $\{\pm 1, \pm i \}$, the perturbation magnitude is $c_{k}=b/(k+1)^{r}$, and the step size actualization of the parameters is given by $a_k= a/(k+1+A)^s$. The performance of the CSPSA algorithm depends on the selection of its gain parameters $A,a,b,r$ and $s$. However, their optimal selection is not clear and depends on the minimized function. Since its proper selection is beyond of a quick analysis, it is recommended to use some well-studied, stable gain combinations. There are two of those, called standard and asymptotic gains which are given by
$A=0, a=3, b=0.1, s=0.602, r = 0.101,$ and $A = 0, a = 3, b = 0.1, s = 1, r = 0.166$, respectively. 
 
In our numerical simulations we use the asymptotic gains, and only the $A$ gain is modified for performance improvements. In stage one, for all the simulations and experiments we use $A = 0$. In stage two, Figs.~\ref{fig:Fig2} and \ref{fig:Fig3} are generated with $A=4$. Figs.~ \ref{fig:Fig1-1}, \ref{fig:Fig1-2}, \ref{fig:Fig1-3} and \ref{fig:Fig1-4} consider $A=32$, $A=16$, $A=8$, and $A=4$ respectively. The noisy simulations from Tables \ref{BP_VDGE} and \ref{BP_iVDGE}, together with the experimental results shown in Figures~ \ref{fig:Fig4-1} and \ref{fig:Fig4-2}, use also the asymptotic gains, with $A=8$ for the second stage of iVDGE.


\end{document}